\documentclass[11pt,a4paper]{article}
\pdfoutput=1
\DeclareMathAlphabet{\mathpzc}{OT1}{pzc}{m}{it}
\usepackage{jheppub}
\usepackage{amsmath,amssymb,amsfonts,array,eucal,mathrsfs,upgreek,stackrel,accents,slashed,graphicx,tensor,color}
\usepackage{hyperref}

\newcommand{\beq}{\begin{equation}}
\newcommand{\beql}[1]{\begin{equation}\label{#1}}
\newcommand{\eeq}{\end{equation}}
\newcommand{\bseql}[1]{\begin{subequations}\label{#1}}
\newcommand{\eseq}{\end{subequations}}
\newcommand{\eq}[1]{\eqref{#1}}

\newcommand{\lf}{\left}
\newcommand{\rf}{\right}
\newcommand{\Quad}{\qquad\qquad}
\newcommand{\tql}[1]{{\textquotedblleft #1\textquotedblright}}
\newcommand{\fr}{\frac}
\newcommand{\rt}{\sqrt}
\newcommand{\ft}{\footnote}
\newcommand{\tn}{\tensor}
\newcommand{\wg}{{\,\wedge\;}}
\newcommand{\ord}{{\boldsymbol{\CMcal{O}}}}
\newcommand{\im}{\Rightarrow}
\newcommand{\iim}{\Leftrightarrow}
\newcommand{\diag}{\mathrm{diag}}
\newcommand{\intdx}[1]{\int\mathrm{d}^{#1}x\,}

\newcommand{\warn}[1]{{\color{red}#1}}
\newcommand{\warnb}[1]{{\color{blue}#1}}

\newcommand{\al}{\alpha}
\newcommand{\ga}{\gamma}
\newcommand{\de}{\delta}
\newcommand{\De}{\Delta}
\newcommand{\ep}{\epsilon}
\newcommand{\ve}{\varepsilon}
\newcommand{\tH}{\theta}
\newcommand{\ka}{\kappa}
\newcommand{\la}{\lambda}
\newcommand{\Sg}{\Sigma}

\newcommand{\cA}{\mathcal{A}}
\newcommand{\cF}{\mathcal{F}}
\newcommand{\tlcF}{\tilde{\mathcal{F}}}

\newcommand{\tlq}{\tilde{q}}

\newcommand{\hatA}{\hat{A}}
\newcommand{\hatB}{\hat{B}}
\newcommand{\hatC}{\hat{C}}
\newcommand{\hatE}{\hat{E}}
\newcommand{\hatf}{\hat{f}}
\newcommand{\hatn}{\hat{n}}
\newcommand{\hatq}{\hat{q}}
\newcommand{\hatr}{\hat{r}}
\newcommand{\hatY}{\hat{Y}}
\newcommand{\hatphi}{\hat{\phi}}

\newcommand{\ed}{\mathrm{d}}
\newcommand{\ds}{\mathrm{d}s}
\newcommand{\dt}{\mathrm{d}t}
\newcommand{\dr}{\mathrm{d}r}
\newcommand{\dhatr}{\mathrm{d}\hat{r}}
\newcommand{\dR}{\mathrm{d}R}
\newcommand{\dw}{\mathrm{d}w}
\newcommand{\dx}{\mathrm{d}x}
\newcommand{\dphi}{\mathrm{d}\phi}
\newcommand{\drho}{\mathrm{d}\rho}
\newcommand{\dl}{\partial}
\newcommand{\Dl}{\nabla}
\newcommand{\LieD}{\pounds}
\newcommand{\met}{\mathsf{g}}
\newcommand{\hmet}{\mathsf{h}}
\newcommand{\Rie}{\mathpzc{R}\,}
\newcommand{\Ric}{\mathpzc{R}\,}
\newcommand{\ExtK}{\mathpzc{K}\,}

\newcommand{\AdS}{\mathbf{AdS}}
\newcommand{\R}{\mathbb{R}}

\newcommand{\GND}[1]{{G^{\scriptscriptstyle{(#1)}}_\text{\sc n}}}
\newcommand{\kaEH}{\kappa_\text{\sc eh}}
\newcommand{\ac}{\CMcal{S}}
\newcommand{\Ham}{\mathsf{H}}
\newcommand{\EM}{{\text{\sc em}}}
\newcommand{\phys}{\mathrm{phys}}
\newcommand{\scLif}{{\text{\tiny Lif}}}
\newcommand{\sch}{\text{sch}}
\newcommand{\ext}{\text{ext}}
\newcommand{\hor}{\text{\sc h}}
\newcommand{\CFT}{\text{\sc cft}}
\newcommand{\LS}{\text{\sc ls}}
\newcommand{\nhLS}{\text{nh-{\sc ls}}}

\definecolor{LightBluejb}{rgb}{0.40,0.50,1.00}

\numberwithin{equation}{section}

\title{Charged Dilatonic AdS Black Branes \\ in Arbitrary Dimensions}

\author{Per Berglund,}
\author{Jishnu Bhattacharyya,}
\author{David Mattingly}
\affiliation{Department of Physics, University of New Hampshire,\\ Durham, NH 03824, USA}
\emailAdd{per.berglund@unh.edu}
\emailAdd{jishnu.b@unh.edu}
\emailAdd{dyo7@cisunix.unh.edu}

\abstract{We study electromagnetically charged dilatonic black brane solutions in arbitrary dimensions with flat transverse spaces, that are asymptotically AdS. This class of solutions includes spacetimes which possess a bulk region where the metric is approximately invariant under Lifshitz scalings. Given fixed asymptotic boundary conditions, we analyze how the behavior of the bulk up to the horizon varies with the charges  and derive the extremality conditions for these spacetimes.}

\begin{document}
\maketitle

\section{Introduction}
Dilatonic black hole/brane solutions in (super)gravity have been studied extensively in the past, see e.g., \cite{Gibbons:1982ih, gibbons-maeda, ghs, shapere-trivedi-wilczek, hs, gregory-harvey, horne-horowitz}. There has also been a good deal of work on dilatonic black holes/branes in asymptotically (anti-)de Sitter spaces; see e.g., \cite{wiltshire-1, wiltshire-2, wiltshire-3, wiltshire-4, Gao:2004tu, Gao:2004tv}. More recently, electromagnetically charged dilatonic black holes/branes have found applications in an offshoot of the AdS/CFT correspondence \cite{adscft:juan, adscft:gkp, adscft:witten-1} (for reviews see \cite{adscft:review:magoo, adscft:review:hor-pol}), the so-called AdS/condensed matter field theory or AdS/CMT correspondence; for some reviews on AdS/CMT, see \cite{adscmt:review:hartnoll-1, adscmt:review:herzog, adscmt:review:mcgreevy}.

The idea that spacetimes in general relativity with a metric of the form
	\beql{lifshitz-spacetimes}
	\ds^2 = -(r/r_0)^{2z}\dt^2 + (r_0/r)^2\dr^2 + (r/r_0)^2\sum_{i = 1}^n(\dx^i)^2
	\eeq
can holographically describe non-relativistic fixed points of quantum field theories was put forward in \cite{lifshitz:kachru-liu-mulligan}; here $r_0$ is a fixed length scale, while the constant $z \geqq 1$, known as the scaling exponent, depends on the non-relativistic fixed point in question.

To appreciate the proposed correspondence, consider a system near a non-relativistic fixed point, with a mass gap $\De$. The energy scale $\De$ of such a system is related to the correlation length $\xi$ through $\De^{-1} \sim \xi^z$. As the critical point is approached, the correlation length diverges, the mass gap vanishes and the system has an effective scale invariance under
	\beql{non-relativistic-scale-transformations}
	x^i \to \la x^i; \qquad r \to \fr{r}{\la}; \qquad t \to \la^z t\,,
	\eeq
with $\la$ being an arbitrary positive constant. Now, the metric \eq{lifshitz-spacetimes} is isometric under the scale transformations \eq{non-relativistic-scale-transformations}, in addition to being trivially invariant under constant translations of $t$ and $x^i$\ft{The metric \eq{lifshitz-spacetimes} does not have boost invariance however. For the construction of holographic duals of quantum field theories with non-relativistic boost invariance in addition to scale invariance of the above type, see \cite{Son:2008ye}, \cite{Balasubramanian:2008dm}.}. In particular, for $z > 1$ we have different scalings of the space and time which is usually the case with non-relativistic fixed points\ft{When $z = 1$ \eq{lifshitz-spacetimes} actually describes the Poincar\'e patch of $(n + 2)$-dimensional AdS ($\AdS_{n + 2}$). We then have additional isometries of the metric corresponding to the Lorentz boosts and the special conformal transformations of the dual relativistic CFT.}. Therefore, based on our general intuition about AdS/CFT, we expect the metric \eq{lifshitz-spacetimes} to holographically capture 
physics of non-relativistic fixed points. Spacetimes with a metric of the form \eq{lifshitz-spacetimes} are called global Lifshitz spacetimes.

Subsequently, with \tql{non-relativistic holography} in mind, global Lifshitz and Lifshitz-Schwarzschild spacetimes as exact solutions of Einstein-Maxwell-dilatonic gravity with negative cosmological constant in four and higher dimensions were found \cite{lifshitz:taylor, lifshitz:gkpt, Chen:2010kn, lifshitz:gkpt-2}. Note that the Lifshitz-Schwarzschild solution has been called \tql{Lifshitz black brane} in the recent literature. We call it the Lifshitz-Schwarzschild solution instead, in order to avoid certain confusions (see below). Our nomenclature is justified in section \ref{L-LBB}.

The gravitational systems considered in \cite{lifshitz:taylor, lifshitz:gkpt, Chen:2010kn}\ft{Note that \cite{lifshitz:taylor} considers other kinds of gravitational systems too.} are very closely related to the one considered in the classic work of \cite{ghs}, in that, they all have an additional negative cosmological constant term in the action. The existence of global Lifshitz and Lifshitz-Schwarzschild solutions in such systems clearly make them candidate holographic duals of non-relativistically scale invariant quantum field theories. However, the presence of the negative cosmological constant in the action indicates that such systems should also posses asymptotically AdS solutions \cite{lifshitz:gkpt, lifshitz:peetetal-1, lifshitz:peetetal-2}\ft{In this work, by AdS we will always mean the Poincar\'e patch of AdS and not the global AdS spacetime.}. Moreover, \cite{lifshitz:gkpt} argued that the near horizon geometry of such an asymptotically AdS black brane, at extremality, should possess Lifshitz like behaviour\ft{This is consistent with our general understanding of charged extremal black hole solutions, e.g., asymptotically flat/AdS extremal Reissner-Nordstr\"om black holes, extremal D3, M2, M5 branes etc. In all these examples the respective Einstein's equations are solved exactly by the corresponding (extremal) near horizon solutions.}. As we will also see, the near horizon geometry of a certain subclass of the general non-extremal asymptotically AdS solutions is Lifshitz-Schwarzschild-like. We refer to such asymptotically AdS solutions with Lifshitz/Lifshitz-Schwarzschild like behaviour near the horizon, as the near horizon Lifshitz black brane and the near horizon Lifshitz-Schwarzschild black brane, respectively. The AdS completions of the near horizon Lifshitz/Lifshitz-Schwarzschild solutions have already been found in \cite{lifshitz:gkpt, lifshitz:peetetal-1, lifshitz:peetetal-2}. Note, in this regard, that the near horizon Lifshitz black brane solution has exact scaling symmetry under \eq{non-relativistic-scale-transformations} near the horizon. Since we call the so-called \tql{Lifshitz black brane} solution as the Lifshitz-Schwarzschild solution in this paper, there should hopefully be no confusion henceforth.

In the current paper, we expand upon the recent work \cite{lifshitz:gkpt, lifshitz:peetetal-1, lifshitz:peetetal-2} on finding asymptotically AdS black brane solutions to Einstein-Maxwell-dilaton gravity with negative cosmological constant in arbitrary dimensions\ft{Certain generalizations of these gravitational systems have also appeared in the literature. Among these,~\cite{Charmousis:2009xr, Charmousis:2010zz, Gouteraux:2011ce} have cosidered replacing the pure cosmological constant term with a (or a sum of) Liouville potential(s), and have also considered even higher dimensional generalizations involving higher rank form fields, such that the previous generalizations appear as Kaluza-Klein reductions of the latter. Another possible generalization involving multiple $U(1)$ fields has been considered in~\cite{Tarrio:2011de}. Note that these references do not necessarily deal with asymptotically AdS solutions.}. There are several reason why such solutions are interesting to study: first of all, as already mentioned above, the gravitational system which admits these solutions are simple generalizations of those studied in \cite{ghs}. It is thus important to investigate the nature of the asymptotically AdS counterparts the asymptotically flat dilatonic black hole solution of \cite{ghs}. It is also useful to compare these solutions with the well known AdS-Reissner-Nordstr\"om black brane solutions, admitted by the same gravitational system when the dilatonic interaction is switched off; we will make some comments on the similarities and the differences among these two classes of solutions below. Finally, since the solutions we are going to discuss are all electrically charged black branes with AdS asymptotics, by the AdS/CFT correspondence they are gravitational duals of (some currently unknown) relativistic CFTs at finite temperature and chemical potential. In particular, the radial evolution of the near horizon Lifshitz black brane solution describes a renormalization group flow between a relativistic conformal fixed point in the UV and a Lifshitz-like fixed point in the IR. In contrast, the near horizon geometry of the $(n + 2)$-dimensional extremal AdS-Reissner-Nordstr\"om black brane solution is $\AdS_2\times\R^n$. Thus, the radial evolution of the extremal AdS-Reissner-Nordstr\"om solution also describes a holographic renormalization group flow between two fixed points, but the IR fixed point in this case is very different from the Lifshitz-like one. In other words, the dilatonic black brane solutions allow us to obtain new kinds of (holographic) renormalization group flows. Embedding the global Lifshitz solution into an asymptotically AdS solution then allows us to interpret the IR fixed point as having an emergent scale invariance just like in the $\AdS_2\times\R^n$ case.

Our central result is the general non-extremal, static, asymptotically AdS black $n$-brane solution, admitted by Einstein-Maxwell-dilatonic gravity with negative cosmological constant in $(n + 2)$-dimensions, with $n \geqq 2$. The solution has a regular horizon and is characterized by three asymptotic charges, namely, the mass, the electric charge and the dilatonic charge. In contrast to their asymptotically flat cousins \cite{ghs}, the dilatonic charge in our case is not a function of the electric charge and the mass, but is a free parameter in itself. We also derive the appropriate extremality conditions satisfied by this general non-extremal solution. Quite interestingly, the extremality condition, which is related to the vanishing of the surface gravity at the horizon, takes the same form as that in \cite{ghs}. These extremality conditions furthermore allow us to identify the appropriate limits in which the general non-extremal solution reduces to the near horizon Lifshitz-Schwarzschild black brane and the near horizon Lifshitz black brane solutions, respectively. In particular, we are able to verify explicitly that the near horizon Lifshitz black brane solution is indeed the extremal solution of the system. To help the reader keep track of the relevant solutions, table \ref{table:list-of-solutions} in section \ref{summary-and-conclusions} provides a summary of all the asymptotically AdS solutions in terms of the appropriate asymptotic charges.

In addition, we study the thermodynamics of these solutions. Since our solutions are asymptotically AdS, such relations also describe the thermodynamics of the dual CFT via holography. Comparing the situation with \cite{ghs}, we notice that unlike the discontinuous drop in the temperature of an asymptotically flat dilatonic black hole in the extremal limit, the temperature of the non-extremal black brane in our case smoothly goes to zero as extremality is approached. Furthermore, the area of the horizon in the case of the extremal (near horizon Lifshitz) solution vanishes, and therefore so does the Bekenstein-Hawking entropy of the brane. This is, however, in stark contrast with the case of the extremal AdS-Reissner-Nordstr\"om (near horizon $\AdS_2\times\R^n$) solution, where the entropy is finite. In other words, although the latter solution can be found from the former in the limit of vanishing dilatonic coupling, even for very small values of the dilatonic coupling the entropy is strictly zero in the former case. This suggests that not all physical quantities (although some are as we are going to show) pertaining to these two cases are smoothly related as the dilatonic coupling is varied.

We conclude the introduction by outlining the rest of the paper: in section \ref{stat-Rniso-met-EOM} we present the action and the equations of motion that we study. We next collect some relevant facts about the global Lifshitz and Lifshitz-Schwarzschild solutions, particularly in the context of our gravitational system, in section \ref{L-LBB}. Our main results including the various solutions and the extremality conditions are contained in sections \ref{asymp-AdS-solns} and \ref{nh-solns}, and in section \ref{thermodynamics} we discuss the thermodynamics of these solutions. We summarize our results and make some final concluding remarks in section \ref{summary-and-conclusions}. We have also included a detailed proof of a no-go result in appendix \ref{V-restriction}: here we show that the only kind of potential for the dilaton allowing an exact, global Lifshitz solution is the negative cosmological constant. Finally, in appendix \ref{appendix:mass} we discuss in some detail the Hawking-Horowitz prescription \cite{haw-hor} to determine the mass of solutions in general relativity, which we have used to find the mass of the various solutions we discuss.
\section{The action and the equations of motion}\label{stat-Rniso-met-EOM}
Following the general remarks above, we want to pursue a  study of $(n + 2)$-dimensional ($n \geqq 2$) static geometries supporting an electromagnetic (two-form) field $\cF_{\mu\nu}$  and a scalar dilaton $\phi$ in the presence of negative cosmological constant\ft{As is shown in appendix \ref{V-restriction} the case of a more general potential $V(\phi)$ for the dilaton reduces to the current situation by demanding that global Lifshitz solutions are admitted.}. A suitable action for this system is\ft{We let $\kaEH^2 = 8\pi\GND{n+2}$, with $\GND{n+2}$ the $(n + 2)$-dimensional Newton's constant. As expected, the physical charges do not depend on this choice.}
	\beql{our-action}
	\ac = \intdx{n + 2}\rt{-\det\met}\lf[\fr{\Rie}{2\kaEH^2} + \fr{n(n+1)}{2\kaEH^2\ell^2} - \fr{1}{2}(\dl\phi)^2 - \fr{e^{2\al\phi}\;\cF^2}{2}\rf] + \text{boundary terms}\,,
	\eeq
where 
	\beq
	\cF^2 = \fr{1}{2}\cF_{\mu\nu}\cF^{\mu\nu}\,.
	\eeq
The boundary terms in the action ensure a well defined variational problem and determine the conserved charges, but the equations of motion are, of course, unaffected by them. For algebraic simplicity, we can absorb the normalization of the Einstein-Hilbert term by making the redefinitions\ft{However, in order to define physical quantities (e.g., mass) associated with the solutions of \eq{our-action} we will need to use the canonically normalized counterparts of the various fields and parameters according to \eq{redefinition}.}:
	\beql{redefinition}
	\phi \to \fr{\phi}{\rt{2\kaEH^2}}; \qquad \al \to \al\rt{2\kaEH^2}; \qquad \cA \to \fr{\cA}{\rt{2\kaEH^2}}\,.
	\eeq
Note that the product $\al\phi$ is unaffected by the redefinition while $\cF_{\mu\nu}$ is redefined in the same way as $\cA_\mu$. In terms of these redefined fields and parameters, the Einstein's equations are 
	\bseql{EOM:const_V}
	\beql{g-EOM:const_V}
	\Ric_{\mu\nu} = -\lf[\fr{n + 1}{\ell^2} + \fr{e^{2\al\phi}\;\cF^2}{2n}\rf]\met_{\mu\nu} + \fr{1}{2}e^{2\al\phi}\;\cF_{\mu\la}\tn{\cF}{_\nu^\la} + \fr{1}{2}(\dl_\mu\phi)(\dl_\nu\phi)\,.
	\eeq
The dilaton's equation of motion is 
	\beql{phi-EOM:const_V}
	\Dl^2\phi = \al\;e^{2\al\phi}\;\cF^2
	\eeq
and the Maxwell's equations are
	\beql{F-EOM:const_V}
	\Dl_\nu(e^{2\al\phi}\;\cF^{\mu\nu}) = 0\,.
	\eeq
	\eseq
In order to find static Lifshitz-like solutions \eq{lifshitz-spacetimes} to the above equations, we want the spacetime to be locally a product of a 2-dimensional space spanned by a global time function $t$ and a radial coordinate $r$, and a flat $\R^n$ along the transverse directions. We also assume that the various fields in the problem can only depend on the radial coordinate. The most general form of a metric with these properties is
	\beql{met:ansatz}
	\ds^2 = -A(r)\dt^2 + C(r)\dr^2 + B(r)\sum_{i = 1}^n(\dx^i)^2
	\eeq
where $i, j = 1, ... , n$ labels the directions along the $\R^n$. The functions $A(r), C(r)$ and $B(r)$ are not independent components of the metric, i.e., we can relate one of them to the other two by a suitable redefinition of $r$. For the dilaton we assume $\phi = \phi(r)$. Finally, we only consider the case when the electromagnetic field is purely electric in nature through the following ansatz
	\beql{F-ansatz}
	\cF_{\mu\nu} = -f_e(r)\rt{A(r)C(r)}\;\ep_{\mu\nu x^1 ... x^n}\,,
	\eeq
where $\ep_{\mu_1\mu_2 \nu_1 ... \nu_n}$ is the completely antisymmetric symbol\ft{The completely antisymmetric (Levi-Civita) tensor $\ve$ is related to $\ep$ through $\ve = \rt{-\det\met}\,\ep$.}, normalized such that $\ep_{tr x^1 ... x^n} = 1$. From the Maxwell's equation \eq{F-EOM:const_V}, we then have
	\beql{F-soln}
	f_e(r) = \fr{q}{\ell}\fr{e^{-2\al\phi(r)}}{B(r)^{\fr{n}{2}}}\,,
	\eeq
where $q$ is a dimensionless constant related to the physical charge. This can be seen as follows: the $n$-form dual field strength of $\cF$ is given by
	\beq
	\tlcF_n = \fr{\ve_{\la_1 ... \la_n \mu\nu}\,e^{2\al\phi}\,\cF^{\mu\nu}}{n!\;2!}\;\dx^{\la_1} \wg ... \wg \dx^{\la_n} = \fr{q}{\ell}\,\dx^1 \wg ... \wg \dx^n\,.
	\eeq
Note that the dual field strength is not the simple Hodge dual but includes a factor of $\exp(2\alpha\phi)$. This ensures that the Maxwell term in the action \eq{our-action} has the canonical form when expressed as the integral of the exterior product of $\cF$ and $\tlcF_n$, i.e.,
	\beq
	\ac_\EM = -\fr{1}{4}\intdx{n + 2}\rt{-\det\met}\;e^{2\al\phi}\cF_{\mu\nu}\cF^{\mu\nu} = -\fr{1}{2}\int\cF\wg\tlcF_n\,.
	\eeq
Now, the physical electric charge density per unit $n$-volume is defined as the (infinite) flux of the canonically normalized (see \eq{redefinition}) counterpart of $\tlcF_{n}$, through a transverse Gaussian surface located at any arbitrary value of the radial coordinate, divided by the (infinite) $n$-volume of the transverse space. This is a good opportunity to point out, that here and later, whenever we come across an \tql{extensive} quantity, i.e., one that depends on the volume of the transverse space, one needs to consider densities per unit $n$-volume. This is because, owing to the infinite volume of the transverse space, the actual extensive quantities are all infinite. With this prescription, the physical electric charge density per unit $n$-volume is related to $q$ through
	\beql{def:physical-electric-charge}
	q_\phys = \fr{q}{\ell\rt{2\kaEH^2}}\,.
	\eeq
Upon these simplifications, the Einstein's equations \eq{g-EOM:const_V} take the following form: the $(t, t)$ components are 
	\bseql{EEq:r}
	\beql{EEq-tt:r}
	\Dl^2\log A(r) = \fr{n-1}{n}\fr{q^2}{\ell^2}\fr{e^{-2\al\phi(r)}}{B(r)^n} + \fr{2(n+1)}{\ell^2}\,.
	\eeq
The non-trivial part of the Einstein's equations along $(i, j)$ is
	\beql{EEq-xx:r}
	\Dl^2\log B(r) = -\fr{1}{n}\fr{q^2}{\ell^2}\fr{e^{-2\al\phi(r)}}{B(r)^n} + \fr{2(n+1)}{\ell^2}\,.
	\eeq
The linear combination of the $(t, t)$ and the $(r, r)$ equations become
	\beql{EEq-tr:r}
	\fr{n}{4}\fr{B'(r)}{B(r)}\lf[\fr{A'(r)}{A(r)} + \fr{C'(r)}{C(r)} - \fr{B'(r)}{B(r)}\rf] - \fr{n}{2}\lf[\fr{\ed}{\dr}\fr{B'(r)}{B(r)}\rf] = \fr{\phi'(r)^2}{2}
	\eeq
and finally, the dilaton equation of motion \eq{phi-EOM:const_V} becomes
	\beql{phiEOM:r}
	\Dl^2\phi(r)  = -\fr{\al q^2}{\ell^2}\fr{e^{-2\al\phi(r)}}{B(r)^n}\,.
	\eeq
	\eseq
Analyzing the solutions to the above equations is the main focus of the current work. In particular, we want to obtain the general, asymptotically AdS non-extremal black brane solution of \eq{EEq:r} and seek its connection to the special cases of asymptotically AdS but near horizon Lifshitz/Lifshitz-Schwarzschild solutions, recently found in \cite{lifshitz:gkpt, lifshitz:peetetal-1, lifshitz:peetetal-2}.
\section{The global Lifshitz and the Lifshitz-Schwarzschild solutions}\label{L-LBB}
Before discussing the asymptotically AdS solutions, we want to briefly review the asymptotically Lifshitz solutions of \eq{EEq:r}, because of their relevance in the near horizon geometry of the special solutions mentioned at the end of the last section.

There exists \cite{lifshitz:taylor, lifshitz:gkpt, Chen:2010kn} a one parameter family of exact solutions of \eq{EEq:r} given as follows:
	\beql{met:LBB}
	A(r) = (r/\ell_\scLif)^{2z}f(r); \qquad C(r) = (r/\ell_\scLif)^{-2}f(r)^{-1}; \qquad B(r) = (r/\ell_\scLif)^2\,,
	\eeq
where
	\beql{f(r):LBB}
	f(r) = 1 - \fr{2m_\LS\ell_\scLif^{n + z}}{r^{n + z}}\,,
	\eeq
with $m_\LS$ being a free parameter (see below), and the dilaton having a logarithmic behaviour
	\beql{phi:LBB}
	\phi(r) = -\fr{n}{\al}\log (r/\ell_\scLif)\,.
	\eeq
The scaling exponent $z$ is given by
	\beql{z:LBB}
	z = 1 + \fr{n}{2\al^2}\,,
	\eeq
while $q = \pm q_\scLif$, where $q_\scLif$ and $\ell_\scLif$ are given by
	\beql{q-scale:LBB}
	q_\scLif = \rt{\fr{2n(n+1)}{2\al^2 + 1}}; \qquad \ell_\scLif = \ell\rt{\fr{(n + z - 1)(n + z)}{n(n + 1)}}\,.
	\eeq 
For $m_\LS = 0$ we recover the scale invariant global Lifshitz solution \eq{lifshitz-spacetimes}. When $m_\LS \neq 0$, we will call this solution the Lifshitz-Schwarzschild solution (for reasons to be explained in a moment). The Lifshitz-Schwarzschild solution has a curvature singularity at $r = 0$ which can be seen from the expressions of the curvature invariants, e.g., the Ricci scalar
	\beq
	\Rie = -\fr{2}{\ell_\scLif^2}\lf\{\lf[\fr{n(n + 1)}{2} + nz + z^2\rf] + \fr{n(z - 1)}{2}\fr{2m_\LS\ell_\scLif^{n + z}}{r^{n+z}}\rf\}\,.
	\eeq
To prevent a naked singularity in the solution, we must therefore insist $m_\LS > 0$ so that there is a horizon at $r=r_\LS$, where $r_\LS$ is given by
	\beql{horizon:LBB}
	r_\LS^{n + z} = 2m_\LS\ell_\scLif^{n + z}\,.
	\eeq
It should also be noted that the curvature invariants are all finite when $m_\LS = 0$\ft{This is true for the invariants Ricci$^2$ and Riemann$^2$ as well. Also, when $z = 1$, the only curvature invariant that is singular when $m_\LS \neq 0$ is Riemann$^2$.}, i.e., the global Lifshitz solution is free of curvature singularities at $r = 0$\ft{As has been observed in \cite{lifshitz:kachru-liu-mulligan, adscmt:review:hartnoll-1}, when $z\neq 1$ the horizon at $r = 0$ does have a pp singularity which renders these spacetimes geodesically incomplete. However it is expected that stringy effects will resolve this mild singularity \cite{hor-ross:naked-bh-1}.}.

The parameter $m_\LS$ is related to the mass per unit $n$-volume of the Lifshitz-Schwarzschild solution (see appendix \ref{appendix:mass})
	\beql{M:LBB}
	M_\LS = \fr{n\,m_\LS}{\kaEH^2\ell_\scLif}\,.
	\eeq
Also, as a black object in general relativity, the Lifshitz-Schwarzschild solution has a temperature associated with the horizon given by
	\beql{temperature:LBB}
	T_\LS = \fr{(n + z)(2m_\LS)^{\fr{z}{n + z}}}{4\pi\ell_\scLif} = \fr{(2m_\LS)^{\fr{z}{n + z}}}{4\pi\ell}\rt{\fr{n(n + 1)(n + z)}{n + z - 1}}\,.
	\eeq
The temperature vanishes in the limit $m_\LS \to 0$ when we also get the global Lifshitz solution back; this is consistent with the scale invariance of the Lifshitz solution. For $m_\LS \neq 0$, but $z = 1$, the expression for the temperature matches that of the AdS-Schwarzschild solution. We also have an entropy density per unit $n$-volume associated with the Lifshitz-Schwarzschild solution, given by the Bekenstein-Hawking formula
	\beql{entropy:LBB}
	S_\LS = \fr{2\pi}{\kaEH^2}\fr{r_\LS^n}{\ell_\scLif^n} = \fr{2\pi(2m_\LS)^{\fr{n}{n + z}}}{\kaEH^2}\,.
	\eeq
From \eq{M:LBB}, \eq{temperature:LBB} and \eq{entropy:LBB} we then have the Smarr's formula \cite{Smarr:1972kt}
	\beql{MTS:LBB}
	M_\LS = \fr{n}{n + z}T_\LS S_\LS\,.
	\eeq		 
For $z = 1$, we get back the corresponding formula for the AdS-Schwarzschild solution. 

Based on our discussion so far, the general Lifshitz-Schwarzschild solution for $z \neq 1$ can be seen to be a straight-forward generalization of the AdS-Schwarzschild solution, just like the global Lifshitz solution (for $z \neq 1$) itself is a generalization of the $\AdS_{n + 2}$ solution. Stated differently, any expression pertaining to these solutions, reduces to the corresponding one pertaining to the $\AdS_{n + 2}$/AdS-Schwarschild solutions, as we set $z = 1$. We are therefore justified in calling the \tql{Lifshitz black brane} solution as the Lifshitz-Schwarzschild solution.

As already mentioned before, these solutions, apart from being exact solutions to \eq{EEq:r}, also arise as near horizon approximations of certain asymptotically AdS solutions of \eq{EEq:r}. It is then natural to ask, as well, about the relationship of these special solutions to the general, non-extremal asymptotically AdS black brane solution of \eq{EEq:r}. We conclude the current section by presenting the Lifshitz and the Lifshitz-Schwarzschild solutions in a way suitable to address these issues, and in sections \ref{extremal-soln} and \ref{nh-LBB-soln} below, we address these in details.

As an artifact of the gauge fixing condition \eq{def:fixed-detg:R} that we impose on the metric in our analysis of the asymptotically AdS solutions, the Lifshitz/Lifshitz-Schwarzschild solutions appear, as near horizon solutions, in a different radial coordinate. If we temporarily denote the new radial coordinate by $\hatr$, then it is related to $r$, appearing in \eq{met:LBB}, through
	\beql{hatr-hatl:LBB-3}
	\hatr(r) = \hat{\ell}_\scLif(r/\ell_\scLif)^{n + z}; \Quad \hat{\ell}_\scLif = \fr{\ell_\scLif}{n + z}\,.
	\eeq
Treating $\hatr$ as a new radial coordinate, the metric \eq{met:LBB} takes the form
	\beql{met:LBB-3}
	\ds^2 = -(\hatr/\hat{\ell}_\scLif)^\ga\hatf(\hatr)\dt^2 + \fr{(\hatr/\hat{\ell}_\scLif)^{-2}}{\hatf(\hatr)}\dhatr^2 + (\hatr/\hat{\ell}_\scLif)^{2\de}\sum_{i = 1}^n(\dx^i)^2; \qquad \hatf(\hatr) = 1 - \fr{2m_\LS\hat{\ell}_\scLif}{\hatr}\,,
	\eeq
where 
	\beql{de-ga:LBB-3}
	\de = \fr{1}{n + z}; \qquad \ga = \fr{2z}{n + z}\,.
	\eeq
In this new coordinate the metric components satisfy $(-\det\met) = \hatA(\hatr)\hatC(\hatr)\hatB(\hatr)^n = 1$. The horizon in this new coordinate is at
	\beq
	\hatr_\LS = 2m_\LS\hat{\ell}_\scLif
	\eeq
and the dilaton is
	\beql{phi:LBB-3}
	\hatphi(\hatr) = -\fr{n}{\al(n + z)}\log(\hatr/\hat{\ell}_\scLif)\,.
	\eeq	
The temperature of the brane, being independent of the choice of coordinate, is still given by \eq{temperature:LBB}. In particular, this is one way to relate the arbitrary constant in $\hatf(\hatr)$ to $2m_\LS$.
\section{Asymptotically AdS solutions}\label{asymp-AdS-solns}
We now turn to the class of solutions of the equations of motion \eq{EEq:r} that admit AdS boundary conditions asymptotically. One can scale out the $\ell$ dependence in \eq{EEq:r} by using a dimensionless radial coordinate $R$, related to $r$\ft{This radial coordinate is different from the one appearing in the Lifshitz-Schwarzschild metric \eq{met:LBB}.} through
	\beql{def:asymp-radial-coord-R}
	R = \fr{r}{\ell}\,.
	\eeq
It is convenient at this stage to fix the redundancy in the metric components which persists so far. We will impose the following 
	\beql{def:fixed-detg:R}
	-\det\met = A(R)C(R)B(R)^n = R^{2n}\,,
	\eeq
so that $C(R)$ is known in terms of $A(R)$ and $B(R)$. The asymptotically AdS boundary conditions now read	
	\beql{asymp-AdS-bndy-cond:R}
	\lim_{R \to \infty}\fr{A(R)}{R^2} = A_0, \qquad \lim_{R \to \infty}\fr{B(R)}{R^2} = B_0^2, \qquad \lim_{R\to\infty}\phi(R) = \phi_\infty\,,
	\eeq
where $A_0$, $B_0$ and $\phi_\infty$ are constants. The gauge choice \eq{def:fixed-detg:R} and the boundary conditions \eq{asymp-AdS-bndy-cond:R} imply that $R^2C(R)$ goes to a constant as $R \to \infty$, which is consistent with asymptotic AdS-ness. Note however, that we have departed from the standard choice of setting $A_0 = B_0 = 1$. Also, as can be seen from the equations of motion \eq{EEq:r}, if $\phi(r)$ is a solution for the dilaton, then so is $\phi(r) + \phi_0$, $\phi_0$ being any constant, provided the $q$ is replaced by $q\exp(\al\phi_0)$. We could have utilized this freedom to set $\phi_\infty$ to zero. As we are going to argue later, using the more general form of boundary condition \eq{asymp-AdS-bndy-cond:R} will help us with our analysis. 
  
Under these assumptions, the equations of motion \eq{EEq:r} take the following form
	\bseql{EEq:fixed-detg:R}
	\beql{EEq-tt:fixed-detg:R}
	\fr{\ed}{\dR}\lf[\fr{A(R)B(R)^n}{R^n}\fr{A'(R)}{A(R)}\rf] = \fr{q(n - 1)}{n}E(R) + 2(n + 1)R^n\,,
	\eeq
	
	\beql{EEq-xx:fixed-detg:R}
	\fr{\ed}{\dR}\lf[\fr{A(R)B(R)^n}{R^n}\fr{B'(R)}{B(R)}\rf] = -\fr{q}{n}E(R) + 2(n + 1)R^n\,,
	\eeq

	\beql{phiEOM:fixed-detg:R}
	\fr{\ed}{\dR}\lf[\fr{A(R)B(R)^n}{R^n}\phi'(R)\rf] = -\al qE(R)\,,
	\eeq

	\beql{EEq-tR:fixed-detg:R}
	n\fr{\ed}{\dR}\lf[\fr{B'(R)}{B(R)}\rf] + \fr{n(n + 1)}{2}\lf[\fr{B'(R)}{B(R)}\rf]^2 - \fr{n^2}{R}\fr{B'(R)}{B(R)} + \phi'(R)^2 = 0\,,
	\eeq
	\eseq
where we have introduced the electrostatic field $E(R)$ as follows
	\beql{def:E(R):fixed-detg:R}
	E(R) \equiv -\cF_{tR}(R) = \fr{q\,e^{-2\al\phi(R)}}{R^{-n}B(R)^n}\,.
	\eeq
	
Let us first review the solutions to \eq{EEq:fixed-detg:R} for the special cases of $q = 0$ (with $\al$ not necessarily zero), and $q \neq 0$, $\al = 0$. 
For the first case, we have the AdS-Schwarzschild black brane solution
	\beql{AdS-Schwarzschild:R}
	A(R) = \fr{R^2}{B_0^{2n}}\lf[1 - \fr{2m}{R^{n + 1}}\rf]; \qquad B(R) = B_0^2R^2; \qquad \phi(R) = \phi_\infty\,,
	\eeq
where $m \geqq 0$ is a constant. We should highlight that the equations of motion (specifically, the normalization of the cosmological constant term in the equations) force the constraint $A_0B_0^{2n} = 1$ upon us, and this is true for all the solutions we have obtained. When $m = 0$ we have the pure $\AdS_{n+2}$ solution; for $m > 0$, the mass per unit $n$-volume $M$ of the AdS-Schwarzschild black brane is related to $m$ (see \eq{appendix:energy-asymp-AdS}) through:  
	\beql{def:Schwarzschild-M}
	M = \fr{n\,m}{\kaEH^2\ell}\,.
	\eeq
The solution has a curvature singularity at $R = 0$, but the singularity is hidden behind a horizon located at $R = R_\sch$ (Schwarzschild radius), where
	\beql{def:Rsch} 
	R_\sch^{n + 1} = 2m\,.
	\eeq
The other case, namely $q \neq 0$, $\al = 0$ leads to the AdS-Reissner-Nordstr\"om black brane solution. Define
	\beql{def:tlq}
	\tlq = \fr{q}{B_0^n\,e^{\al\phi_\infty}}
	\eeq
in terms of which the AdS-Reissner-Nordstr\"om solution takes the form	
	\beql{AdS-RN:R}
	A(R) = \fr{R^2}{B_0^{2n}}\lf[1 - \fr{2m}{R^{n+1}} + \fr{\tlq^2}{2n(n-1)R^{2n}}\rf]; \qquad B(R) = B_0^2R^2; \qquad \phi(R) = \phi_\infty\,.
	\eeq
The parameter $m$ has the same interpretation \eq{def:Schwarzschild-M} as that in the AdS-Schwarzschild solution (\eq{AdS-RN:R} reduces to \eq{AdS-Schwarzschild:R} as $q \to 0$). A horizon (and hence a physical solution to hide the curvature singularity at $R = 0$) exists only if the extremality condition $m \geqq m_\ext$ is met, where
	\beql{AdS-RN:extremality}
	m_\ext = \lf(\fr{n}{n - 1}\rf)R_{\hor, \ext}^{n+1}; \qquad |q| = \rt{2n(n+1)}\,R_{\hor, \ext}^n\,.
	\eeq

For the more general case of $q \neq 0$, $\al \neq 0$ the above equations \eq{EEq:fixed-detg:R} cannot be solved in a closed form analytically. In the following, we will instead present a solution to \eq{EEq:fixed-detg:R} as a power series in $R^{-1}$. The solution accurate up to $\ord(R^{-(2n+2)})$ is
	\bseql{asymp-soln:fixed-detg:R}
	\beql{asymp-soln:A:fixed-detg:R}
	A(R) = \fr{R^2}{B_0^{2n}}\lf[1 - \fr{2m}{R^{n+1}} + \fr{\tlq^2}{2n(n-1)R^{2n}} - \fr{2\al^2\mu_\phi^2}{nR^{2n+2}} + \ord\lf(R^{-(2n+3)}\rf)\rf]\,,
	\eeq

	\beql{asymp-soln:B:fixed-detg:R}
	B(R) = B_0^2R^2\lf[1 - \fr{2\al^2\mu_\phi^2}{nR^{2n+2}} + \ord\lf(R^{-(2n+3)}\rf)\rf]\,,
	\eeq

	\beql{asymp-soln:phi:fixed-detg:R}
	\phi(R) = \phi_\infty + \fr{2\al\mu_\phi}{R^{n+1}} - \fr{\al\tlq^2}{2n(n-1)R^{2n}} + \fr{2\al m\mu_\phi}{R^{2n+2}} + \ord\lf(R^{-(2n+3)}\rf)\,.
	\eeq
	\eseq
This is the \tql{large $R$ form} of the asymptotically AdS dilatonic black brane solution we are after in the present work. 

A few comments about this form of the solution are in order: first, the parameter $m$  has the same interpretation as above, i.e., the energy density per unit $n$-volume of the brane is still given by \eq{def:Schwarzschild-M}. Secondly, apart from $m$ and $q$, the solution is characterized by a third constant $\mu_\phi$, which appears in \eq{asymp-soln:fixed-detg:R} as a constant of integration, and is proportional (up to unimportant constants) to the dilatonic charge of the brane\ft{This can be shown, for instance, by comparing the first subleading term ($\ord(R^{-(n + 1)})$ in our case) in the asymptotic expansion of the dilaton \eq{asymp-soln:phi:fixed-detg:R} to that of \cite{ghs}.}. From the $\al$ dependence of the above solution (at least up to the order shown), it can be seen that in the limit $\al \to 0$, the solution reduces to the AdS-Reissner-Nordstr\"om solution. As we will show in the next section, $q = 0 \iim \mu_\phi = 0$, so that when $q = 0$ (even if $\al \neq 0$), the solution reduces to the AdS-Schwarzschild solution. From the asymptotic analysis it can be concluded that \eq{asymp-soln:fixed-detg:R} is the unique, static, asymptotically AdS solution to \eq{EEq:fixed-detg:R}.
\section{Near horizon analysis}\label{nh-solns}
By computing the curvature invariants (in powers of $R^{-1}$), it can be seen that \eq{asymp-soln:fixed-detg:R} has a curvature singularity at $R = 0$. Therefore, there must be an event horizon at a positive value of $R$ which hides the singularity. We will denote the radial location of the outermost event horizon by $R_\hor$. Being static, the system admits a timelike Killing vector $\chi_t = \{1, 0, ..., 0\}$\ft{The normalization of the Killing vector at infinity is irrelevant for our present argument.}, and by the usual requirement of the vanishing of the norm of $\chi_t$ on a regular horizon, we have $A(R_\hor) = 0$.

To study the nature of a solution near the horizon, it is necessary to expand the various functions as a power series in $(r - r_\hor)$. The analysis is facilitated when done in terms of the dimensionless radial coordinate $w$ defined as
	\beql{def:nh-radial-coord-w}
	w = \fr{r - r_\hor}{r_\hor} = \fr{R - R_\hor}{R_\hor} \qquad\iim\qquad R = R_\hor(1 + w)\,.
	\eeq
We include in our definition the restriction $w \geqq 0$, such that the coordinate $w$ is well suited to study only the spacetime outside the horizon. Manifestly, $w = 0$ is where the horizon is located, and by \tql{near horizon} we mean small $w$, i.e., $0 < w \ll 1$ (equivalently $R_\hor < R \ll 2R_\hor$).

The gauge fixing condition for the metric \eq{def:fixed-detg:R} now reads 
	\beql{def:fixed-detg:w}
	A(w)C(w)B(w)^n = R_\hor^{2n}(1 + w)^{2n}\,.
	\eeq
We now need to make appropriate series expansion ans\"atze for the various functions, to be valid in the near horizon region. Since $A(w)$ must vanish on the horizon, we make the following ansatz (factors of $R_\hor$ have been included in the following expressions for future convenience)
	\beql{nh-A:fixed-detg}
	A(w) = a_0R_\hor^2w^\ga\hatA(w); \qquad \hatA(w) = 1 + a_1 w + a_2 w^2 + \ord(w^3)\,,
	\eeq
where $a_0 > 0$ and $\ga \geqq 1$, with $\ga = 1$ (at least) for the non-extremal case. For the case of $B(w)$ we make a similar ansatz
	\beql{nh-B:fixed-detg}
	B(w) = b_0^2R_\hor^2w^{2\de}\hatB(w); \qquad \hatB(w) = 1 + b_1w + b_2w^2 + \ord(w^3)\,,
	\eeq
where $b_0 > 0$ and $\de \geqq 0$, so that $B(w)$ is allowed to vanish on the horizon. When we solve the equations of motion \eq{EEq:fixed-detg:R} starting from the horizon, we need to specify boundary conditions on the horizon. The possible vanishing of $A(w)$ and $B(w)$ are already captured through their postulated dependence on $w^\ga$ and $w^{2\de}$, respectively. The constants $a_0$ and $b_0$ in \eq{nh-A:fixed-detg} and \eq{nh-B:fixed-detg} respectively, similar to $A_0$ and $B_0$ in \eq{asymp-AdS-bndy-cond:R}, are to be specified as boundary conditions on the horizon, and cannot be solved through the equations of motion.

When the preceding ansatz for $B(w)$ is used in \eq{EEq-tR:fixed-detg:R} we find the need to allow for a logarithmic divergence in $\phi(w)$ near the horizon; in other words, we need the following ansatz for $\phi(w)$  
	\beql{nh-phi:fixed-detg}
	\phi(w) = \phi_0\log w + \phi_c + \hatphi(w); \qquad \hatphi(w) = \phi_1w + \phi_2w^2 + \ord(w^3)\,,
	\eeq
where $\phi_0$ and $\phi_c$ are constants. Plugging in the above ansatz for $\phi(w)$ in \eq{EEq-tR:fixed-detg:R} we obtain
	\beql{EEq-tw:fixed-detg:w}
	\begin{split}
	& \fr{2n\de\{\de(n + 1) - 1\} + \phi_0^2}{w^2} + \fr{2n\de}{w}\lf[(n + 1)\fr{\hatB'(w)}{\hatB(w)} - \fr{n}{1 + w}\rf] \\
	& \Quad + n\fr{\ed}{\dw}\lf[\fr{\hatB'(w)}{\hatB(w)}\rf] + \fr{n(n + 1)}{2}\lf[\fr{\hatB'(w)}{\hatB(w)}\rf]^2 - \fr{n^2}{1 + w}\lf[\fr{\hatB'(w)}{\hatB(w)}\rf] + \hatphi'(w)^2 = 0\,.
	\end{split}
	\eeq
The lowest order term in the equation imposes the following relation between $\de$ and $\phi_0$
	\beql{reln:de-phi0:fixed-detg}
	2n\de\lf[\de(n + 1) - 1\rf] + \phi_0^2 = 0\,.
	\eeq
In particular, if $\de$ vanishes (as is expected for the non-extremal solution) then so does $\phi_0$ and vice-versa\ft{There is another case when $\phi_0$ must vanish, namely when $\de = (n + 1)^{-1}$. As we are going to argue, this can only be true when $q = 0$, that is, for the trivial case of the AdS-Schwarzschild solution.}. We can now express the equations of motion \eq{EEq-tt:fixed-detg:R} through \eq{phiEOM:fixed-detg:R} entirely in terms of the functions $\hatA(w), \hatB(w)$ and $\hatphi(w)$ as follows: 
	\bseql{EEq:fixed-detg:w}
	\beql{EEq-tt:fixed-detg:w}
	k\fr{\ed}{\dw}\lf[w^{\ga + 2n\de - 1}\hatY(w)\lf[\ga + w\fr{\hatA'(w)}{\hatA(w)}\rf]\rf] {=} \fr{\hatq^2(n - 1)}{n}w^{-2(\al\phi_0 + n\de)}\hatE(w) + 2(n + 1)(1 + w)^n,
	\eeq
	
	\beql{EEq-xx:fixed-detg:w}
	k\fr{\ed}{\dw}\lf[w^{\ga + 2n\de - 1}\hatY(w)\lf[2\de + w\fr{\hatB'(w)}{\hatB(w)}\rf]\rf] = -\fr{\hatq^2}{n}w^{-2(\al\phi_0 + n\de)}\hatE(w) + 2(n + 1)(1 + w)^n,
	\eeq

	\beql{phiEOM:fixed-detg:w}
	k\fr{\ed}{\dw}\lf[w^{\ga + 2n\de - 1}\hatY(w)\lf[\phi_0 + w\hatphi'(w)\rf]\rf] = -\al\hatq^2w^{-2(\al\phi_0 + n\de)}\hatE(w),
	\eeq
	\eseq
where
	\beql{def:k-hatq:fixed-detg}
	k = a_0b_0^{2n}; \qquad \hatq = \fr{q\,e^{-\al\phi_c}}{b_0^nR_\hor^n} \quad\iim\quad q = \hat{q}e^{\al\phi_c}b_0^nR_\hor^n
	\eeq
and
	\beql{def:hatE-hatY:fixed-detg}
	\hatE(w) = \fr{e^{-2\al\hatphi(w)}(1 + w)^n}{\hatB(w)^n}; \qquad \hatY(w) = \fr{\hatA(w)\hatB(w)^n}{(1 + w)^n}\,.
	\eeq
Clearly we have the same number of unfixed \tql{boundary parameters}, namely $B_0$ and $\phi_\infty$ on the asymptotic boundary versus $b_0$ and $\phi_c$ on the horizon (by \eq{def:k-hatq:fixed-detg} the value of $a_0$ is fixed in terms of $b_0$). Whichever side we start from, choosing the values for one pair fixes the values for the other. We will leave this \tql{normalization freedom} manifest in all the expressions to follow, except in those cases where we make an explicit choice.

The linear combination formed by adding $(1/\al)$ times \eq{phiEOM:fixed-detg:w} to the difference of \eq{EEq-tt:fixed-detg:w} and \eq{EEq-xx:fixed-detg:w} takes the following form
	\beql{Eqcombo-txphi:w}
	k\fr{\ed}{\dw}\lf[w^{\ga + 2n\de - 1}\hatY(w)\lf[\lf[\ga - 2\de + \fr{\phi_0}{\al}\rf] + w\lf[\fr{\hatA'(w)}{\hatA(w)} - \fr{\hatB'(w)}{\hatB(w)} + \fr{\hatphi'(w)}{\al}\rf]\rf]\rf] = 0\,.
	\eeq
When \eq{Eqcombo-txphi:w} is considered to the lowest order, we face two possibilities: either $\ga = 1 - 2n\de$ or $\ga \neq 1 - 2n\de$. In the following section \ref{non-extremal-soln} we will consider the former case, and then in section \ref{extremal-soln} we will separately analyse the latter possibility.
\subsection{The non-extremal solution}\label{non-extremal-soln}
The first case we consider is when $\ga = 1 - 2n\de$. As we will show below, this choice gives the non-extremal solution. Expanding the equations of motion \eq{EEq:fixed-detg:w} to the first subleading order, we further  
impose $\phi_0 = -n\de/\al$ since otherwise 
$\hatq = 0$. But $\hatq = 0$ corresponds to the AdS-Schwarzschild black brane solution. From \eq{reln:de-phi0:fixed-detg} we then find either $\de = 0$ or $\de = (n + z)^{-1}$ where $z$ is the scaling exponent as in \eq{z:LBB}. In particular, this rules out the possibility $\phi_0 = 0$, $\de = (n + 1)^{-1}$.
Furthermore, given the near horizon ans\"atze for $A(w)$ \eq{nh-A:fixed-detg} and $B(w)$ \eq{nh-B:fixed-detg}, the near horizon behaviour of the expansion of radial null geodesics is \cite{gr:books}
	\beql{expansion:nh:fixed-detg}
	\ell\tH(w) = \fr{na_0b_0^nR_\hor^{n+1}}{2}w^{-n\de}\lf[1 + \ord(w)\rf]\,,
	\eeq
where we have used that $\ga = 1 - 2n\de$. The expansion $\tH(w)$ should vanish on a true horizon, but if we choose $\de\neq0$ for the present case, we see it actually diverges. Therefore the appropriate choices for the exponents are
	\beq
	\de = \phi_0 = 0; \qquad \ga = 1\,.
	\eeq
In other words, $A(w)$ vanishes linearly on the horizon while $B(w)$ and $\phi(w)$ are regular on the horizon. By defining a new near horizon radial coordinate $\rho = 2\ell\rt{w/k}$, the near horizon metric can now be brought to the Rindler form $\ds^2 = -\ka^2\rho^2\dt^2 + \drho^2 + \ds_\perp^2$, where $\ds_\perp^2$ is the metric on the transverse space and
	\beql{non-ext:surface-gravity} 
	\ka = \fr{kR_\hor}{2b_0^n\ell}
	\eeq	
is the surface gravity. From the Rindler form of the metric, we note $\dl N/\dl \rho = \ka$, where $N = \ka\rho$ is the lapse function. Therefore, the horizon is a regular horizon \cite{gr:books}, and the solution thus describes a non-extremal black brane.

The coefficients to determine the functions $\hatA(w)$ \eq{nh-A:fixed-detg}, $\hatB(w)$ \eq{nh-B:fixed-detg} and $\hatphi(w)$ \eq{nh-phi:fixed-detg} accurately up to $\ord(w^2)$ are given as follows:
	\beql{coeffs:nh:non-ext:ord-1}
	a_1 = \lf[\fr{(2n - 1)\hatq^2 - 2n(n^2 - 1)}{2nk} + \fr{n}{2}\rf]; \quad b_1 = -\lf[\fr{\hatq^2 - 2n(n + 1)}{nk}\rf]; \quad \phi_1 = -\fr{\al\hatq^2}{k}\,,
	\eeq

	\begin{align}\label{coeffs:nh:non-ext:ord-2}
   \nonumber a_2 & = \fr{2\{(3n - 2)\al^2 + 9n - 5\}\hatq^4 - 8n(3n^2 + n - 2)\hatq^2 + 12n^2(n + 1)^2(n - 1)}{12nk^2} \\
       \nonumber & \quad + \fr{(2n - 1)\hatq^2 - 2n(n^2 - 1)}{2k} + \fr{n(n - 1)}{6}\,, \\
             b_2 & = -\lf[\fr{(2n\al^2 + n - 1)\hatq^4 - 4n(n^2 - 1)\hatq^2 + 4n^2(n + 1)^2(n - 1)}{4n^2k^2} + \fr{\hatq^2 - 2n(n + 1)}{2k}\rf]\,, \\
\nonumber \phi_2 & = -\lf[\fr{\al(2n\al^2 + n + 1)\hatq^4 - 2n\al(n + 1)(2n + 1)\hatq^2}{4nk^2} + \fr{n\hatq^2\al}{2k}\rf]\,.
	\end{align}

To be able to connect the near horizon solution found above to the asymptotic solution \eq{asymp-soln:fixed-detg:R}, we need to relate the parameters $k$ and $\hatq$ characterizing near horizon form of the solution to $m$, $\mu_\phi$ and $q$ characterizing the asymptotic form. Our goal in the remainder of this section will be to accomplish that. To that end, consider integrating the equations \eq{EEq-tt:fixed-detg:R}, \eq{EEq-xx:fixed-detg:R} and \eq{phiEOM:fixed-detg:R} from $R = R_\hor$ to $R = \infty$. The analysis is facilitated by the following observation: from the expression of $E(R)$ \eq{def:E(R):fixed-detg:R} the electrostatic potential at the horizon, $\Phi_\hor$, is 
	\beql{def:Phih:fixed-detg:R}
	\Phi_\hor = -\int_\infty^{R_\hor}\dR\;E(R) = q\int_{R_\hor}^\infty\dR\;\fr{e^{-2\al\phi(R)}}{R^{-n}B(R)^n}\,.
	\eeq
Note that $(q^{-1}\Phi_\hor) > 0$ since the integrand is positive\ft{Work done to bring a unit charge (of the same type the black brane carries) to the horizon is positive.}. Upon performing the above mentioned integration\ft{The $R^n$ terms on the right hand side of \eq{EEq-tt:fixed-detg:R} and \eq{EEq-xx:fixed-detg:R} are divergent when the upper limit of the integration is taken to $\infty$. There are however identical divergent pieces on the left hand side of both the equations coming from $A(R) \sim R^2$ and $B(R) \sim R^2$ when $R$ is large. Thus, to make things well defined one should first integrate up to  $R = R_\text{large}$, cancel out the potentially divergent pieces from both sides of the equations and finally take $R_\text{large} \to \infty$.} and rearranging the equations a bit, we obtain
	\beql{reln:after-integration}
	2m + \fr{2 - k}{n - 1}R_\hor^{n+1} = \fr{q\Phi_\hor}{n}; \qquad 2m - R_\hor^{n+1} = \fr{q\Phi_\hor}{2n}; \qquad 2(n + 1)\mu_\phi = q\Phi_\hor\,.
	\eeq
These are three equations in three unknowns: $R_\hor, k$ and $\Phi_\hor$. The last equation in fact gives $\Phi_\hor$ in terms of $\mu_\phi$ and $q$. Since $q^2(q^{-1}\Phi_\hor) > 0$, we can further conclude 
	\beql{positivity-mu-phi:fixed-detg}
	\mu_\phi \geqq 0
	\eeq
and the equality holds iff $q = 0$ (i.e., only for the AdS-Schwarzschild solution). Solving the remaining equations, we find  
	\beql{Rh:non-ext:fixed-detg}
	 R_\hor^{n+1} = 2m - \lf(\fr{n + 1}{n}\rf)\mu_\phi = 2m - \fr{q\Phi_\hor}{2n} = R_\sch^{n+1}\lf[1 - \lf(\fr{n + 1}{2n}\rf)\fr{\mu_\phi}{m}\rf]\,,
	\eeq
where $R_\sch^{n + 1} = 2m$ \eq{def:Rsch}, and
	\beql{k:non-ext:fixed-detg}
	k = \fr{2(n + 1)(m - \mu_\phi)}{R_\hor^{n + 1}} = (n + 1)\lf[\fr{1 - \displaystyle{\fr{\mu_\phi}{m}}}{1 - \displaystyle{\lf(\fr{n + 1}{2n}\rf)\fr{\mu_\phi}{m}}}\rf]\,.
	\eeq
Since $k > 0$ by definition \eq{def:k-hatq:fixed-detg}, we must have $m > \mu_\phi$ for the non-extremal case. Also, for a fixed $m$, $\mu_\phi$ can increase until it equals $m$ at which point $k$ vanishes. Therefore, 
	\beql{extremality-condition-1}
	m \geqq \mu_\phi
	\eeq
represents one of the extremality conditions.
It is very interesting to note that the above condition is identical to the equivalent one in \cite{ghs}. When we discuss the thermodynamics of this solution in section \ref{thermodynamics}, we will show that the above condition, when saturated, is equivalent to the vanishing of the surface gravity at the horizon. This is then consistent with the expectations of an extremality condition.

\begin{figure}[t!]
	\centering
	\includegraphics[scale=0.9]{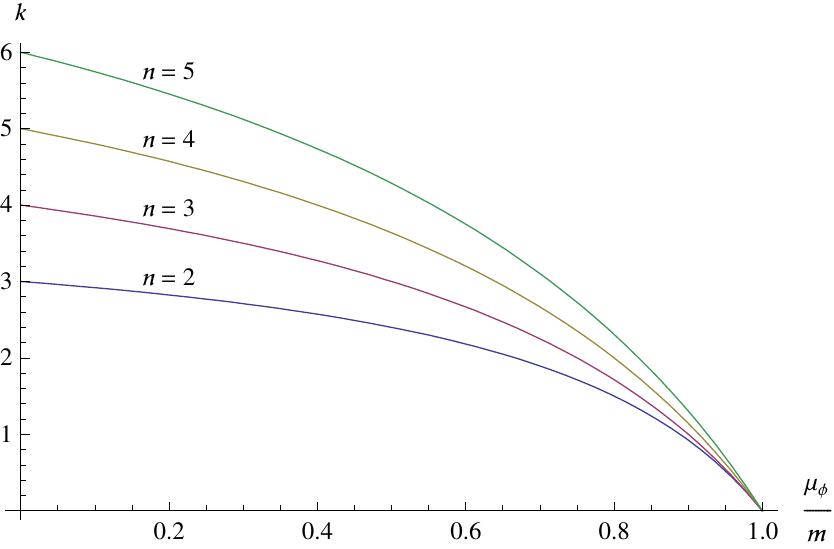}
	\includegraphics[scale=0.9]{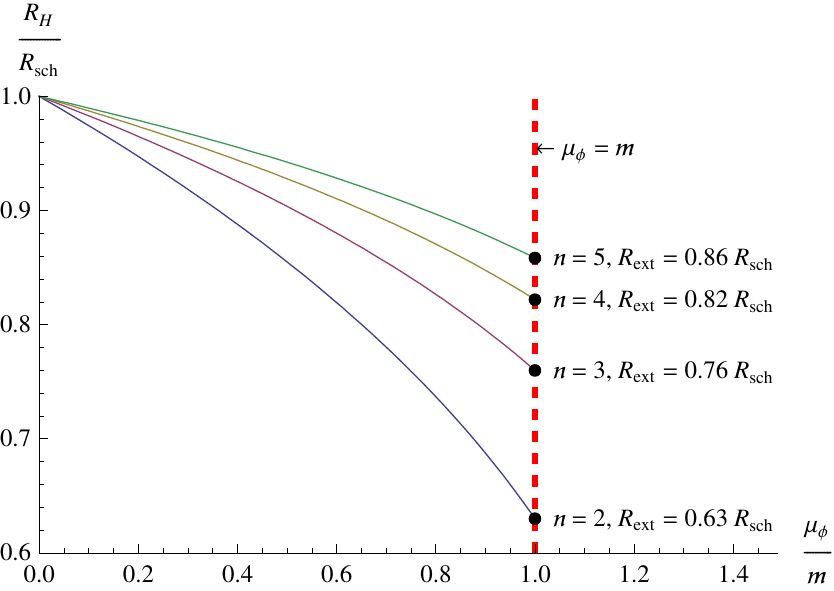}
	\caption{$k$ (left) and $\fr{R_\hor}{R_\sch}$ (right) plotted as functions of $\fr{\mu_\phi}{m}$ for $n = 2$ to 5. The parameter $k$ starts at $(n + 1)$ when $\mu_\phi = 0$ and goes to zero as $\mu_\phi \to m$. Correspondingly, the ratio $\fr{R_\hor}{R_\sch}$ starts at one for all $n$ at $\mu_\phi = 0$ and reaches the extremal value of the ratio when $m = \mu_\phi$ (dashed vertical line).}
	\label{figure:k-and-Rh-vs-x}
	\end{figure}

The bounds \eq{positivity-mu-phi:fixed-detg} and \eq{extremality-condition-1} on $\mu_\phi$ are equivalent to the following bounds on $k$
	\beql{k:non-ext:bound}
	0 \leqq k \leqq (n + 1)\,.
	\eeq
The upper-bound on $k$ is true when $\mu_\phi = 0$, i.e., when $q = 0$. We then have the AdS-Schwarzschild solution \eq{AdS-Schwarzschild:R} and it can be easily checked that $A(w) = (n + 1)wR_\sch^2 + \ord(w^2)$ in this case. The left plot in figure \ref{figure:k-and-Rh-vs-x} shows the behaviour of $k$ as a function of $\mu_\phi/m$.

There seems to be an apparent mismatch in the number of parameters labeling a given solution. From the near horizon side these parameters are (na\"ively) $k$ and $\hatq$, whereas from the asymptotic side they are $m$, $q$ and $\mu_\phi$. This, as we are going to argue now, is not conflicting. The black hole solutions we describe include a static matter field (the dilaton) outside the horizon. It is thus natural to expect that the location of the horizon should not be solely determined by the total mass and the total electric charge carried by the spacetime, but also by the configuration of the dilaton in the bulk. In other words, the location of the horizon, $R_\hor$, should be an independent parameter that needs to be specified in addition to $m$ and $q$ to completely describe the solution\ft{It then seems to be more of a coincidence, that the dilatonic charge is determined by the total electromagnetic charge and the mass of an asymptotically flat dilatonic black holes \cite{ghs}, and in spite of the presence of a non-trivial dilaton in the bulk, the location of the horizon is still determined just by these two parameters.}. From \eq{k:non-ext:fixed-detg}, \eq{Rh:non-ext:fixed-detg} and \eq{def:k-hatq:fixed-detg} this is equivalent to choosing $m$, $q$ and $\mu_\phi$. We discuss this issue in more details at the end of section \ref{nh-LBB-soln}.

On the other hand, starting from near the horizon, the relevant parameters to choose are $k$, $\hatq$, and $m$. It can be shown from \eq{EEq:fixed-detg:R} that two solutions with the same values for the ratios $\dfrac{\mu_\phi}{m}$ and $\dfrac{q}{m^{\fr{n}{n+1}}}$ are related to each other by a transformation involving a constant scaling of the radial coordinate. This is also naturally contained in the fact that for the class of solutions with the same values for the parameters $k$ and $\hatq$
	\beq
	\fr{\mu_\phi}{m} = \fr{2n(n + 1 - k)}{(n + 1)(2n - k)}; \qquad \fr{q}{(2m)^{\fr{n}{n+1}}} = \hatq\lf[\fr{n - 1}{2n -k}\rf]^{\fr{n}{n + 1}}\,,
	\eeq
which follows from \eq{k:non-ext:fixed-detg}, \eq{Rh:non-ext:fixed-detg} and \eq{def:k-hatq:fixed-detg}. The infinite class of such solutions are however distinguishable by the fact that they all have different values for $R_\hor$ \eq{Rh:non-ext:fixed-detg} and thus, even though the solutions are same as functions of $w$, they are different as functions of $R$.
	\begin{figure}[h!]
	\centering
	\includegraphics[scale=0.6]{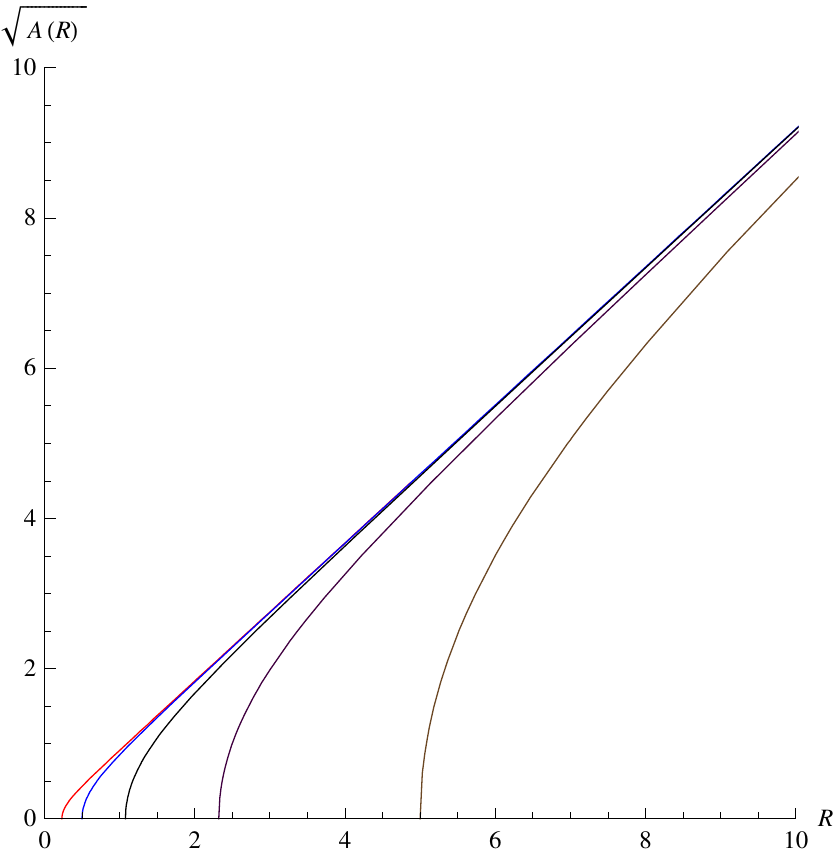}
	\includegraphics[scale=0.6]{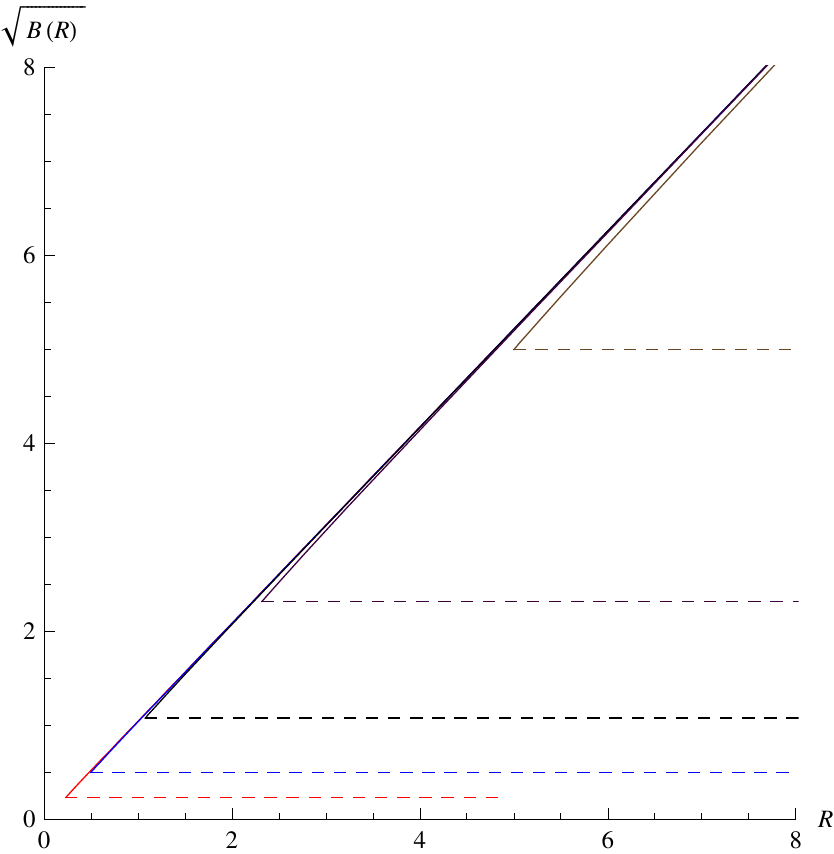}
	\includegraphics[scale=0.6]{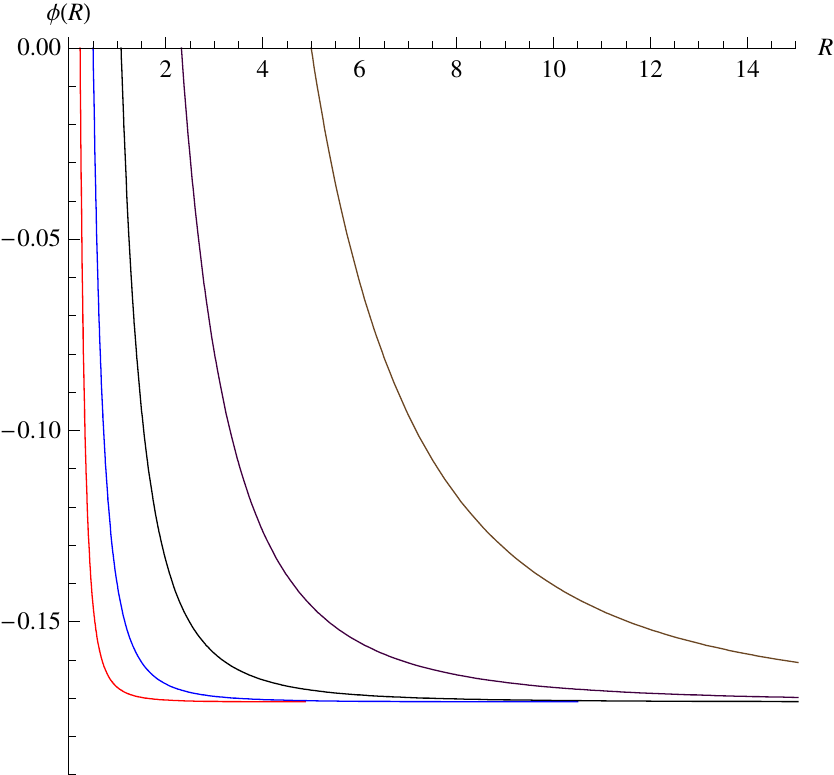}
	\caption{$\rt{A(R)}, \rt{B(R)}$ and $\phi(R)$ for $n = 2$, $\al = 1$, $k = 2.4$ ($\mu_\phi = 0.5\,m$) and $\hatq = 0.5\,q_\scLif$. The curves for $\rt{A(R)}$ and $\phi(R)$, from left to right, correspond to $m = 0.01$, $0.1$, $1.0$, $10$ and $100$, respectively. The curves for $\rt{B(R)}$ follow the same colour pattern, and for each mass, $\rt{B(R)}$ ends at the corresponding value of $R_\hor$ shown by the same coloured dotted line. The square roots of $A(R)$ and $B(R)$ were plotted to show their linear (asymptotic AdS) nature as $R$ becomes large. The coefficients $A_0$ and $B_0$ \eq{asymp-AdS-bndy-cond:R} are both very close to one in all the cases shown in the figure.}
	\label{figure:non-ext-asymp}
	\end{figure}  

So far we have only dealt with the approximate forms of the solution near the horizon and in the asymptotic region. To find the solution in the bulk of the spacetime we numerically integrated the equations \eq{EEq:fixed-detg:w} to solve for the functions $\hatA(w)$, $\hatB(w)$ and $\hatphi(w)$. We used the {\tt NDSolve} routine of {\sc Mathematica} to perform the numerical integration. Stating from slightly outisde the horizon ($w_\text{min} \sim 10^{-17}$), we set the boundary conditions on $\hatA(w)$, $\hatB(w)$ and $\hatphi(w)$ by evaluating the functions and their derivatives at this point using the near horizon coefficients \eq{coeffs:nh:non-ext:ord-1} and \eq{coeffs:nh:non-ext:ord-2} and ran the integration up to $w_\text{max} \sim 10^{10}$.

To relate the hatted functions to the actual metric coefficients and the dilaton through \eq{nh-A:fixed-detg}, \eq{nh-B:fixed-detg} and \eq{nh-phi:fixed-detg}, we need to pick values for $b_0$ and $\phi_c$--we made the most natural choice of setting $b_0 = 1$, $\phi_c = 0$. 
In figure \ref{figure:non-ext-asymp} we show some of our numerical results, with the caption of the figure explaining the details. Based on our earlier remarks on relating the near horizon parameters $k$ and $\hatq$ with the asymptotic ones i.e., $m$, $\mu_\phi$ and $q$, we could also check that the asymptotic forms of the functions agree very well with the corresponding numerically evauated functions in the large $w$ region. 

Although the numerical results shown in figure \ref{figure:non-ext-asymp} were for a specific value of $n$ and $\al$, we found similar results for other values of these parameters as well. In \eq{k:non-ext:bound} we already found upper and lower bounds on the values of the parameter $k$; this naturally narrows down the allowed parameter space to explore. The bound \eq{k:non-ext:bound} on $k$, as noted earlier, is equivalent to the extremality bound \eq{extremality-condition-1}. However, $\mu_\phi$ being independent of $m$ and $q$ in the present situation, we do not have a bound on $q$. This, as we already pointed out, is contrary to  
the usual charged black hole/brane solutions, and particularly the asymptotically flat dilatonic black hole solution \cite{ghs}. There the dilatonic charge is related to the total mass and the electromagnetic charge, so that the extremality condition implies an upper bound on the total electromagnetic charge. In the following two sections, we will find that there actually exists a bound on the parameter $\hatq$ \eq{def:k-hatq:fixed-detg} (although not on $q$ itself) and identify the value of this bound.
\subsection{The extremal solution}\label{extremal-soln}
The other case to consider is when $\ga \neq 1 - 2n\de$, which as we will see corresponds to the extremal case. Analyzing the equations of motion, \eq{EEq:fixed-detg:w} and \eq{Eqcombo-txphi:w} (to lowest order), we find the complete lowest order solution\ft{We put a subscript \tql{$\ext$} on the quantities pertaining to this case, to distinguish them from the corresponding ones in the non-extremal case.}
	\beql{nh-Lifshitz}
	A_\ext(w) = \fr{k_\ext R_{\hor, \ext}^2}{b_{0, \ext}^{2n}}w^\ga; \quad B_\ext(w) = b_{0, \ext}^2R_{\hor, \ext}^2w^{2\de}; \quad \phi_\ext(w) = \phi_{c, \ext} + \phi_0\log w
	\eeq
where the various exponents and coefficients (as defined in \eq{nh-A:fixed-detg}, \eq{nh-B:fixed-detg} and \eq{nh-phi:fixed-detg}) take the values
	\beql{de-ga-phi0:ext:fixed-detg}
	\de = \fr{1}{n + z}; \qquad \ga = \fr{2z}{n + z}; \qquad \phi_0 = -\fr{n}{\al(n + z)}\,,
	\eeq
with $z$, the scaling exponent \eq{z:LBB}, given by
	\beq
	z = 1 + \fr{n}{2\al^2}\,.
	\eeq
while the constants $b_{0, \ext}$ and $\phi_{c, \ext}$ stay unconstrained. Furthermore, the constants $k$ and $\hatq$, defined in \eq{def:k-hatq:fixed-detg}, are also fixed by \eq{EEq:fixed-detg:w} and \eq{Eqcombo-txphi:w}, and take the following values
	\beql{kext-qext:fixed-detg}
	k_\ext = \fr{n(n + 1)(n + z)}{n + z - 1} = \fr{\ell^2}{\hat{\ell}_\scLif^2}; \qquad |\hatq_\ext| = q_\scLif = \rt{\fr{2n(n + 1)}{1 + 2\al^2}}\,,
	\eeq
where $q_\scLif$ and $\hat{\ell}_\scLif$ were defined in \eq{q-scale:LBB} and \eq{hatr-hatl:LBB-3}, respectively. It is easy to check that \eq{nh-Lifshitz}  solves the equations of motion \eq{EEq:fixed-detg:w} and \eq{EEq-tw:fixed-detg:w} exactly. Furthermore,  \eq{nh-Lifshitz} has the same form as the global Lifshitz solution \eq{met:LBB-3} (when $m_\LS = 0$). In fact, these two metrics are identical if we set $b_{0, \ext} = k_\ext^{\de/2}$ and $\phi_{c, \ext} = \phi_0\log\rt{k_\ext}$ as boundary conditions on the horizon, identify $w\rt{k_\ext}$ with $\hatr/\hat{\ell}_\scLif$ in \eq{hatr-hatl:LBB-3} and rescale both $t$ and $x^i$ in \eq{met:LBB-3} by $R_{\hor, \ext}$\ft{Rescaling of $t$ and $x^i$ however affects the lapse and the measure on the transverse space; we will discuss the consequences of this when we consider the thermodynamics of our solutions.}.  

The near horizon solution satisfies $A(w)C(w)B(w)^n = R_{\hor, \ext}^{2n}$ (as opposed to a one on the right hand side because of the rescaling of $t, x^i$ -- compare with \eq{met:LBB-3}). This certainly agrees with the gauge fixing condition \eq{def:fixed-detg:w} near the horizon, but the Lifshitz solution is an exact solution everywhere only if the right hand side of \eq{def:fixed-detg:w} equals $R_{\hor, \ext}^{2n}$ everywhere. This is a virtue of our gauge fixing condition \eq{def:fixed-detg:w}, in that, it distinguishes between the solution which is Lifshitz all the way out, from the one which is Lifshitz only near the horizon.

The change in the behaviour of the solution from Lifshitz to AdS is effected by the correction functions $\hatA_\ext(w)$, $\hatB_\ext(w)$ and $\hat\phi_\ext(w)$ as defined in \eq{nh-A:fixed-detg}, \eq{nh-B:fixed-detg} and \eq{nh-phi:fixed-detg}. The series expansion coefficients of these functions about $w = 0$, up to $\ord(w^2)$, are found by solving \eq{EEq:fixed-detg:w} to the same order and are given as follows
	\beql{coeffs:nh:ext:ord-1}
	a_{1, \ext} = \fr{n(n + 2\al^2)}{n + 2(n + 1)\al^2}; \quad b_{1, \ext} = \fr{2n\al^2}{n + 2(n + 1)\al^2}; \quad \phi_{1, \ext} = -\fr {n^2\al}{n + 2(n + 1)\al^2}\,,
	\eeq
	\begin{align}\label{coeffs:nh:ext:ord-2}
   \nonumber a_{2, \ext} & = \fr{n(n + 2\al^2)(n(7n - 4) + 2(n - 1)(4 + n)\al^2)}{12(n + 2(n + 1)\al^2)^2}\,, \\
	     b_{2, \ext} & = \fr{n\al^2(n(n - 4) + 2(n - 1)(n + 4)\al^2)}{6(n + 2(n + 1)\al^2)^2}\,, \\
\nonumber \phi_{2, \ext} & = -\fr {(n - 4)n^2\al}{12(n + 2(n + 1)\al^2)}\,.
	\end{align}
To connect the near horizon solution for this case to the asymptotic solution \eq{asymp-soln:fixed-detg:R} we integrate \eq{EEq-tt:fixed-detg:R}, \eq{EEq-xx:fixed-detg:R} and \eq{phiEOM:fixed-detg:R}, as before, from $R = R_{\hor, \ext}$ to $R = \infty$. Defining $\Phi_{\hor, \ext}$ analogous to \eq{def:Phih:fixed-detg:R} we again obtain a set of three equations, which are however nothing but those in \eq{reln:after-integration} with $k = 0$\ft{Warning: the constant $k$ \eq{k:non-ext:fixed-detg} does not reduce to $k_\ext$ in the extremal limit $\mu_\phi \to m$, but vanishes.}. Solving these we obtain, first
	\beql{extremality-mu_phi}
	\mu_{\phi, \ext} = m\,.
	\eeq
This shows that the extremality bound \eq{extremality-condition-1} is saturated by this solution. One should also note that this particular solution exists only for a special value of the parameter $\hatq$, i.e., when $|\hatq_\ext| = q_\scLif$ \eq{kext-qext:fixed-detg}. This is very similar to general extremal black hole/brane solutions, which can also carry only a specific amount of charge. In the following section, we will argue that the absolute value of $\hatq$ for a non-extremal solution is bounded from above precisely by $q_\scLif$. The present solution therefore saturates both the extremality bounds, prompting us to call it the extremal solution.

Next, from integrating \eq{EEq-tt:fixed-detg:R}, \eq{EEq-xx:fixed-detg:R} and \eq{phiEOM:fixed-detg:R} from $R = R_{\hor, \ext}$ to $R = \infty$ we also find
	\beql{Rh:ext:fixed-detg}
	R_{\hor, \ext}^{n+1} = \lf(\fr{n - 1}{n}\rf)m\,,
	\eeq
which is the $\mu_\phi \to m$ limit of \eq{Rh:non-ext:fixed-detg}. This is also to the analogous relation \eq{AdS-RN:extremality} for the extremal AdS-Reissner-Nordstr\"om black brane. Finally, similar to \eq{reln:after-integration}, we get
	\beq
	q_\ext\Phi_{\hor, \ext} = 2(n + 1)\mu_{\phi, \ext} = 2(n + 1)m\,,
	\eeq
where $q_\ext$ is related to $\hatq_\ext$ according to \eq{def:k-hatq:fixed-detg} (since $R_\hor = R_{\hor, \ext}$ for the present case), and the last equality follows from the extremality bound \eq{extremality-mu_phi}.

One can check that the near horizon Lifshitz solution reduces to the $\AdS_2\times\R^n$ near horizon limit of the extremal AdS-Reissner-Nordstr\"om solution when $\al \to 0$\ft{The other special limit, namely $\al \to \infty$, does give the pure AdS solution, but in an unusual radial coordinate.}, which also happens to be an exact solution to \eq{EEq:r} when $\al = 0$. This is consistent with the fact that the whole system reduces to the AdS-Reissner-Nordstr\"om solution as $\al \to 0$.
	\begin{figure}[h!]
	\centering
	\includegraphics[scale=0.6]{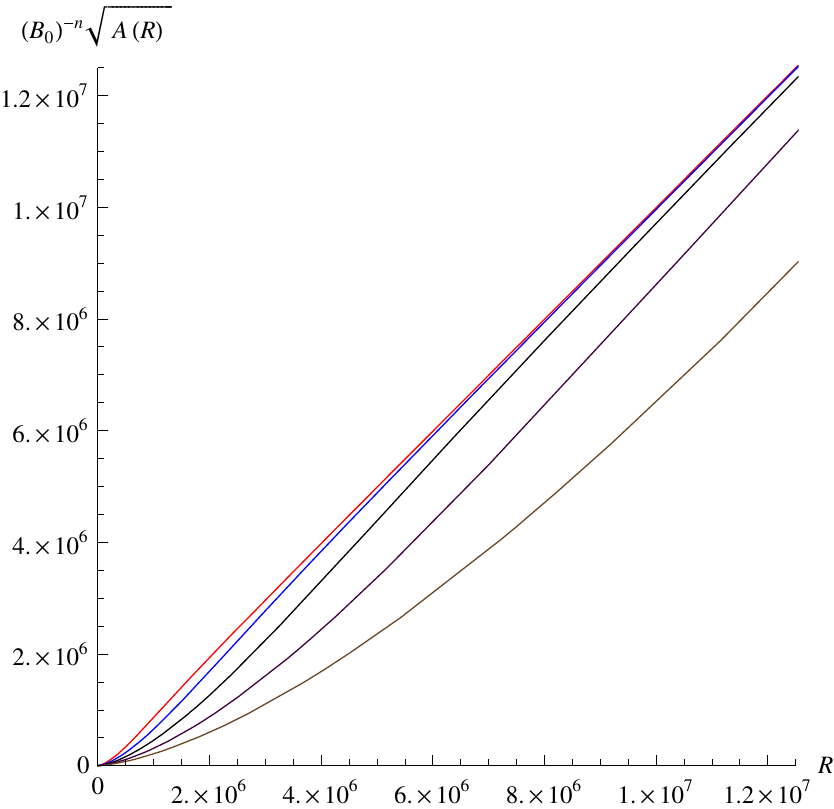}
	\includegraphics[scale=0.6]{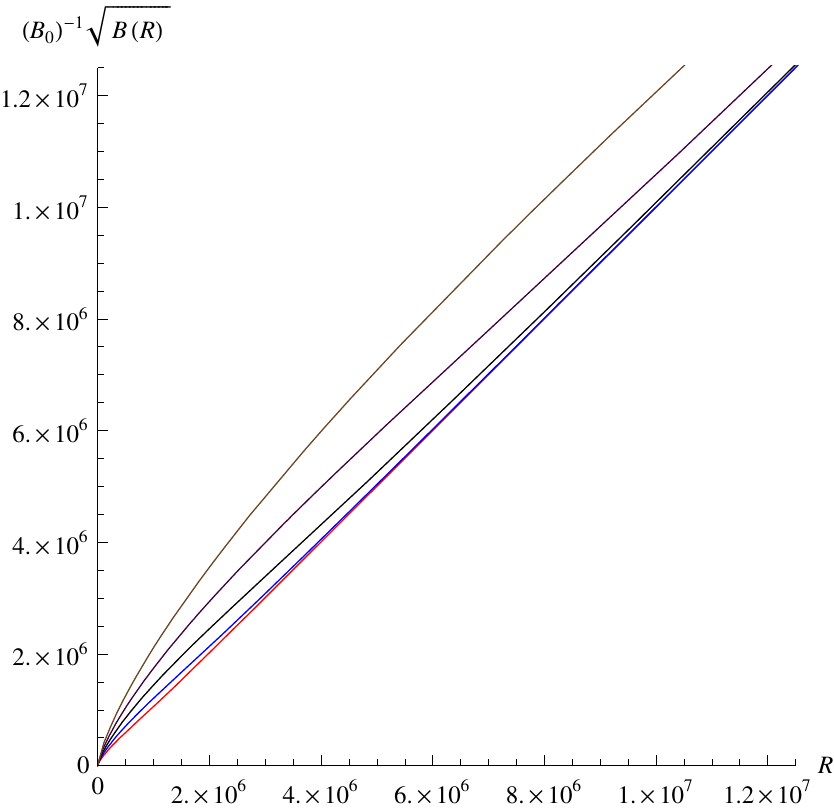}
	\includegraphics[scale=0.6]{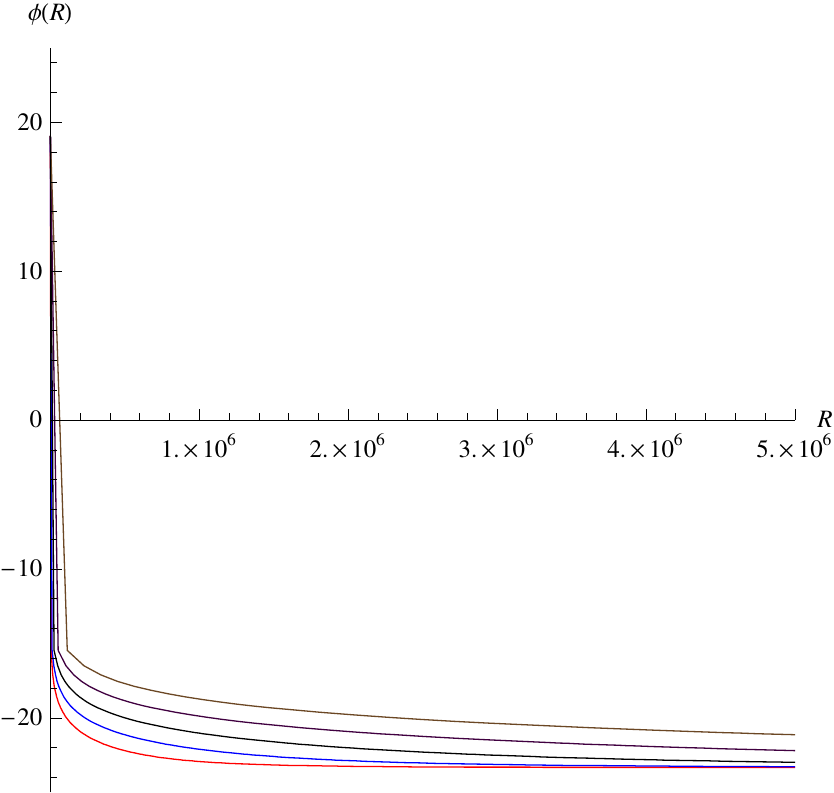}
	\caption{The functions $B_0^{-n}\rt{A(R)}, B_0^{-1}\rt{B(R)}$ and $\phi(R)$ in the bulk for $n = 2$ and $\al = 1$. The plots for $B_0^{-n}\rt{A(R)}$ and $\phi(R)$, from left to right, correspond to $m = 0.01$, $0.1$, $1.0$, $10$ and $100$, respectively, while the curves for $B_0^{-1}\rt{B(R)}$ follow the same colour pattern. The functions $A(R)$ and $B(R)$ were divided by the normalization constants $A_0$ and $B_0^2$ \eq{asymp-AdS-bndy-cond:R} and their square roots were plotted to show their linear (asymptotic AdS) nature as $R$ becomes large. Compared with the non-extremal case (figure \ref{figure:non-ext-asymp}), the functions asymptote to AdS much slowly.}
	\label{figure:ext-asymp}
	\end{figure}

\begin{figure}[h!]
	\centering
	\includegraphics[scale=0.6]{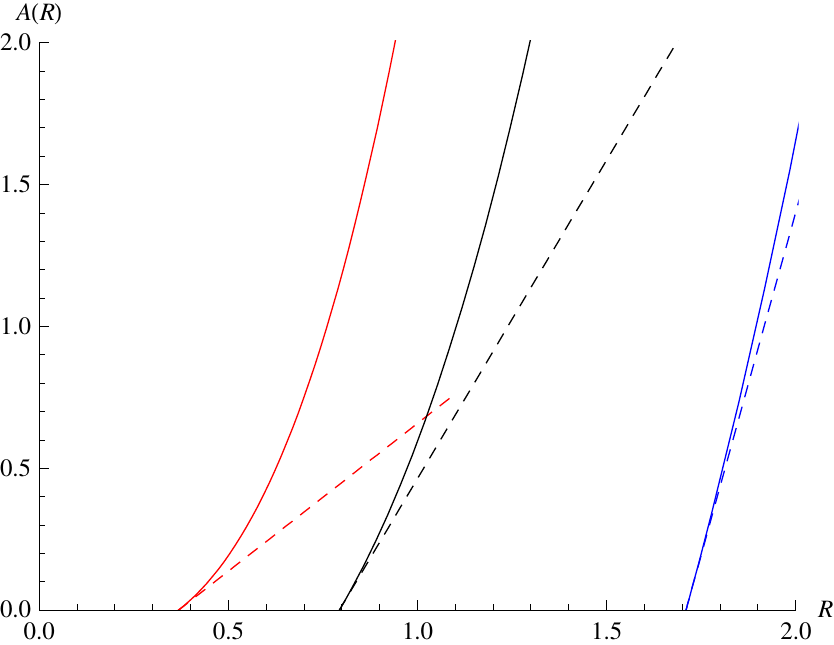}
	\includegraphics[scale=0.6]{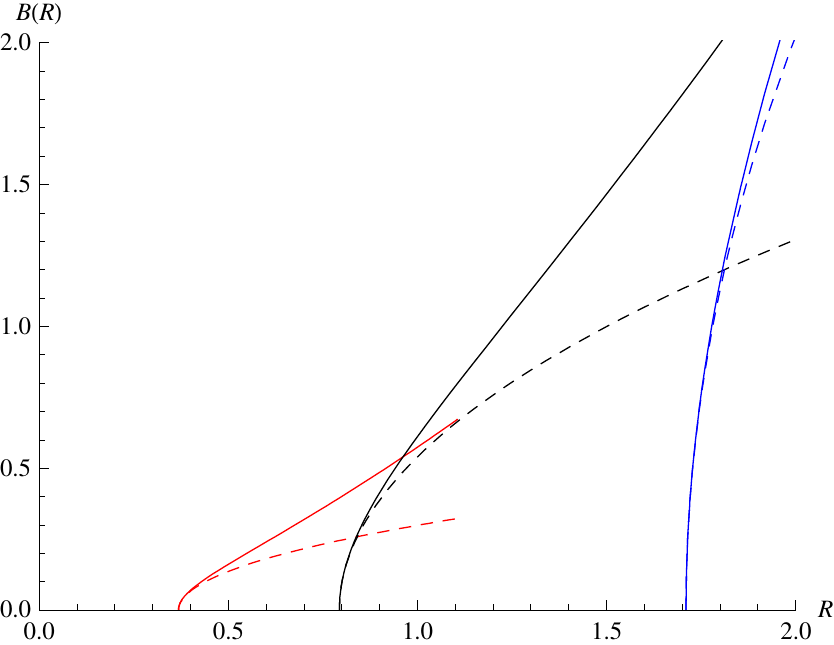}
	\includegraphics[scale=0.6]{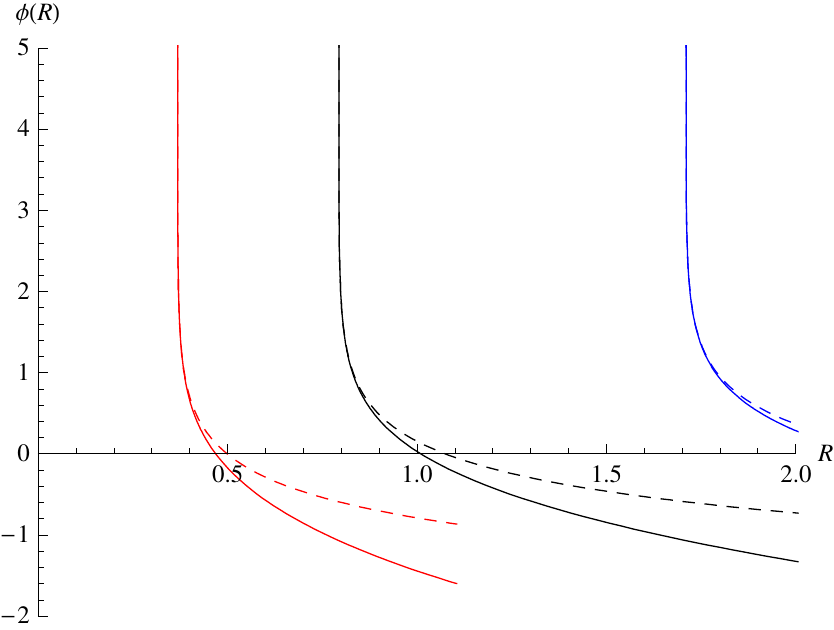}
	\caption{Comparing $A(w), B(w)$ and $\phi(w)$ near the horizon (solid) with the corresponding global Lifshitz solutions (dotted). We have set $n = 2$ and $\al = 1$ ($q_\scLif = 2$). The three plots in each sub-figure correspond to \warn{$m = 0.1$}, $m = 1.0$ and \warnb{$m = 10$}, respectively, with the corresponding values \warn{$R_{\hor, \ext} = 0.37$}, $R_{\hor, \ext} = 0.79$ and \warnb{$R_{\hor, \ext} = 1.7$} for the horizon radii.}
	\label{figure:ext-nh}
	\end{figure}

In the bulk of the spacetime, the nature of the solution can be obtained by numerically integrating\ft{The numerical solution in the bulk for this case has been previously obtained in \cite{lifshitz:gkpt}, \cite{lifshitz:peetetal-1} and \cite{lifshitz:peetetal-2}.} the equations \eq{EEq:fixed-detg:w}. The procedure is similar to that performed for the non-extremal solution. As boundary conditions on the horizon, we set $b_{0, \ext} = k_\ext^{\de/2}$ and $\phi_{c, \ext} = \phi_0\log\rt{k_\ext}$, so that, as discussed above, the Lifshitz nature of the solution is most clear. In figure \ref{figure:ext-asymp} we show that the solutions, for various allowed values of the parameters, are all asymptotically AdS while in figure \ref{figure:ext-nh} we compare the near horizon behaviour of the same set of solutions to the corresponding global Lifshitz solutions. 
\subsection{The near horizon Lifshitz-Schwarzschild solution}\label{nh-LBB-soln}
A natural question to ask, is, where does the Lifshitz-Schwarzschild solution (the $m_\LS \neq 0$ case of \eq{met:LBB-3}) fits in this set-up. Indeed, one finds that the following ansatz solves \eq{EEq:fixed-detg:w} and \eq{EEq-tw:fixed-detg:w} exactly \cite{lifshitz:taylor, lifshitz:gkpt, Chen:2010kn}
	\beql{nh-LBB}
	A(w) = \fr{k_\ext R_\hor^2}{b_0^{2n}}w(w + w_0)^{\ga - 1}; \, B(w) = b_0^2R_\hor^2(w + w_0)^{2\de}; \, \phi(w) = \phi_c + \phi_0\log(w + w_0)\,,
	\eeq
where $\de$, $\ga$, $\phi_0$ and $k_\ext$ are as in the extremal solution, \eq{de-ga-phi0:ext:fixed-detg} and \eq{kext-qext:fixed-detg}, while $\phi_c$ and $b_0$ are unconstrained as before. It also follows from analyzing \eq{EEq:fixed-detg:w} that the solution \eq{nh-LBB} has the same value $|\hatq| = q_\scLif$ as the extremal brane \eq{kext-qext:fixed-detg}. The constant $w_0$ in \eq{nh-LBB} is a free parameter such that $w_0 = 0$ corresponds to the near horizon Lifshitz solution \eq{nh-Lifshitz}. Since $A(w)$ in \eq{nh-LBB} vanishes linearly in $w$, this solution must correspond to a special case\ft{Since $|\hatq|$ is not arbitrary but takes a particular value.} of the non-extremal brane solution studied in section \ref{non-extremal-soln}.

Our first task is to show that this solution is indeed the Lifshitz-Schwarschild solution. As with the near horizon Lifshitz solution, this can be done most easily by choosing the boundary conditions $b_0 = k_\ext^{\de/2}$ and $\phi_c = \phi_0\log\rt{k_\ext}$ on the horizon, rescale $t$ and $x^i$ to absorb the factors of $R_\hor$ and identify $(w + w_0)\rt{k_\ext}$ with the Lifshitz radial coordinate in \eq{hatr-hatl:LBB-3}. This then allows us to identify $w_0$ with the mass parameter of the Lifshitz-Schwarzschild \eq{met:LBB-3} through
	\beql{w0-mLif:LBB-3}
	w_0 = \fr{2m_\LS}{\rt{k_\ext}} = \fr{2m_\LS\hat{\ell}_\scLif}{\ell}\,,
	\eeq
with the last equality following from \eq{kext-qext:fixed-detg}. We should stress that the above identification depends on the particular boundary condition we choose on the horizon. 

On the other hand, to find out the corrections that take the near horizon Lifshitz-Schwarzschild solution to AdS asymptotically, we introduce the correction functions $\hatA_\nhLS(w)$, $\hatB_\nhLS(w)$ and $\hatphi_\nhLS(w)$, in terms of which the asymptotically AdS function can be expressed as
	\bseql{LBB-asymp-AdS}
	\beq
	A_\nhLS(w) = \fr{k_\nhLS R_\hor^2\,w}{b_{0, \nhLS}^{2n}}\lf[\lf(1 + \fr{w}{w_0}\rf)^{\ga - 1} + \hatA_\nhLS(w)\rf]\,,
	\eeq

	\beq
	B_\nhLS(w) = b_{0, \nhLS}^2R_\hor^2\lf[\lf(1 + \fr{w}{w_0}\rf)^{2\de} + \hatB_\nhLS(w)\rf]\,,
	\eeq

	\beq	
	\phi_\nhLS(w) = \phi_{c, \nhLS} + \lf[\phi_0\log\lf(1 + \fr{w}{w_0}\rf) + \hatphi_\nhLS(w)\rf]\,,
	\eeq	
	\eseq
where the subscript \tql{$\nhLS$} stands for near horizon Lifshitz-Schwarzschild. In \eq{LBB-asymp-AdS} above, we have also reverted back to a general boundary condition on the horizon (i.e., arbitrary $b_0$ and $\phi_c$) and introduced
	\beql{k:LBB-3}
	b_{0, \nhLS} = b_0w_0^\de; \qquad k_\nhLS = k_\ext w_0; \qquad \phi_{c, \nhLS} = \phi_c + \phi_0\log w_0\,.
	\eeq
Comparing with \eq{nh-A:fixed-detg}, \eq{nh-B:fixed-detg} and \eq{nh-phi:fixed-detg}, it is clear that the solution is nothing but a non-extremal solution (with a specific value of $|\hatq|$). In particular, we need to choose the correction functions $\hatA_\nhLS(w)$, $\hatB_\nhLS(w)$ and $\hatphi_\nhLS(w)$ such that the leading order terms in all of them vanish linearly on the horizon. Then the complete functions in the big square brackets in \eq{LBB-asymp-AdS} are precisely the hatted functions for the non-extremal case, with the coefficients of series expansion in $w$ being those in \eq{coeffs:nh:non-ext:ord-1} and \eq{coeffs:nh:non-ext:ord-2} up to $\ord(w^2)$.
	\begin{figure}[h!]
	\centering
	\includegraphics[scale=0.6]{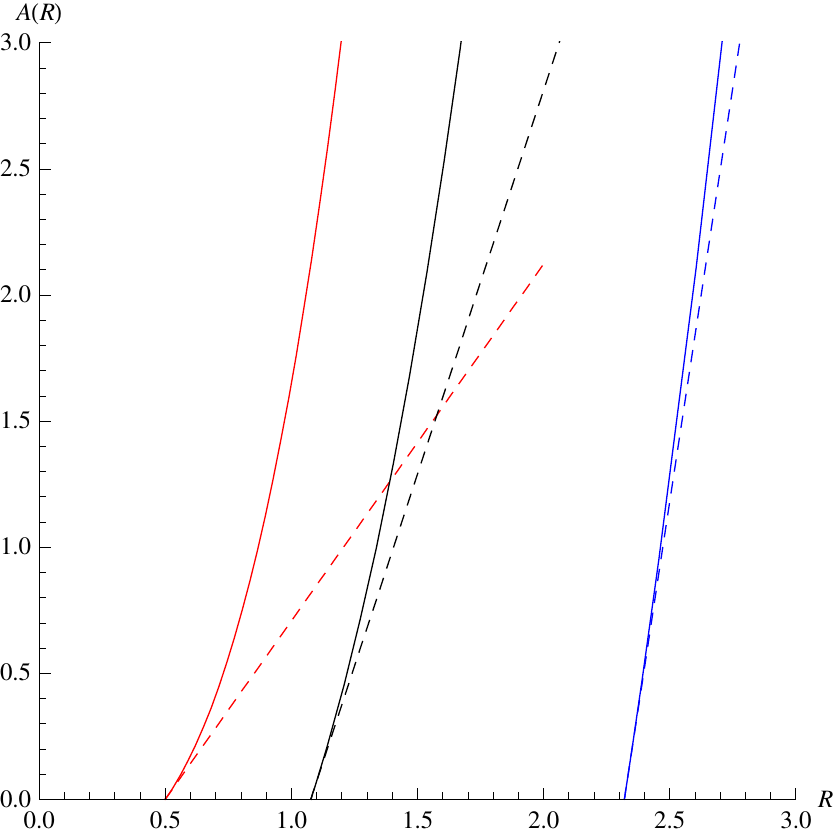}
	\includegraphics[scale=0.6]{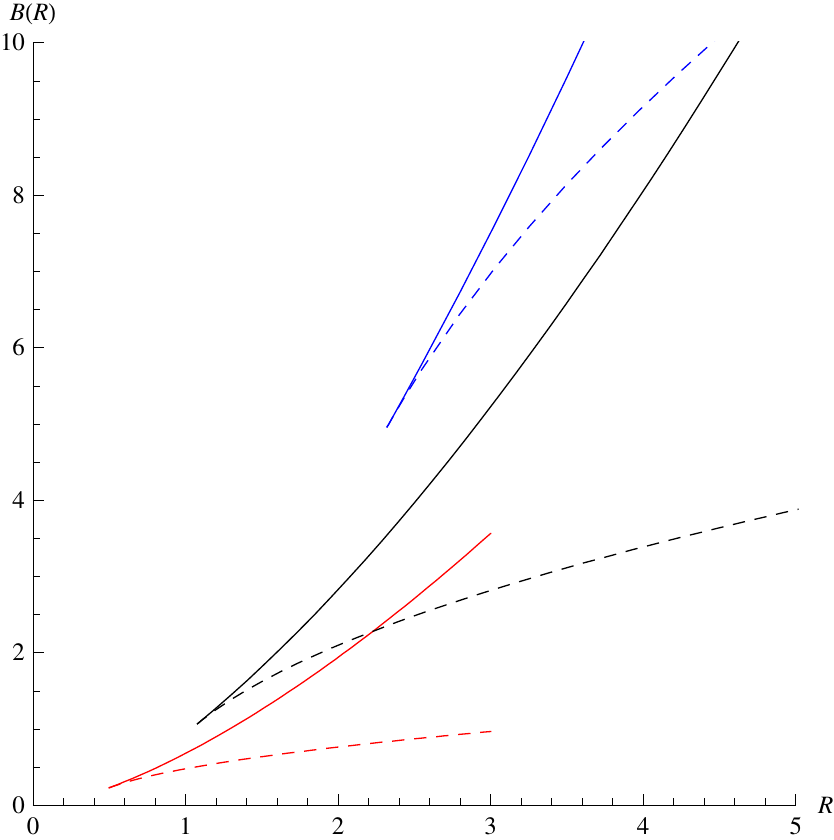}
	\includegraphics[scale=0.6]{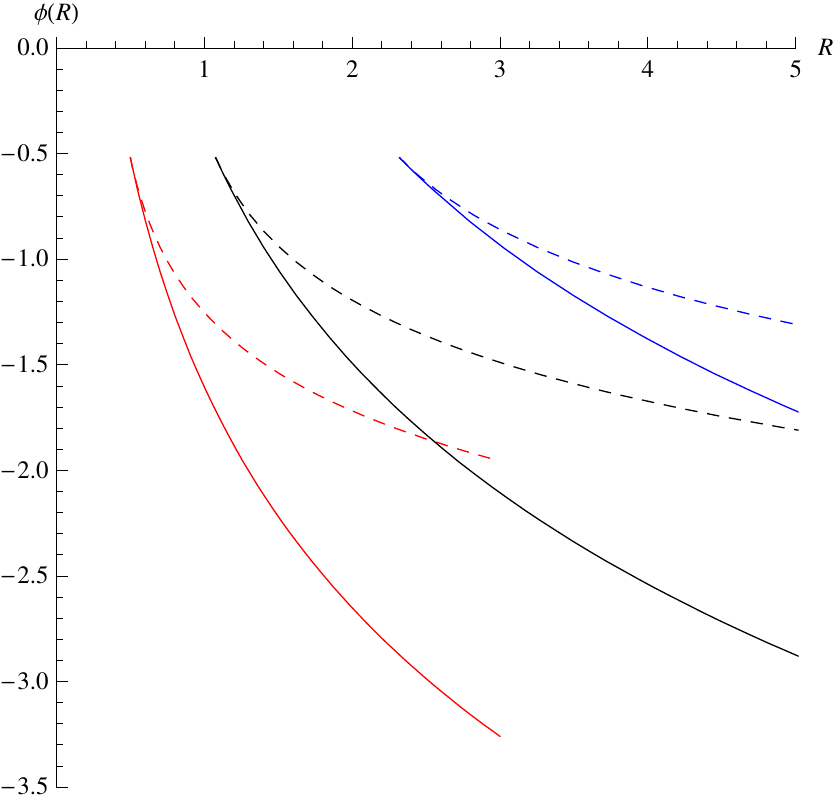}
	\caption{Comparing $A(w), B(w)$ and $\phi(w)$ near the horizon (solid) with the corresponding Lifshitz-Schwarzschild solutions (dotted). We have set $n = 2$, $\al = 1$ ($q_\scLif = 2$) and have chosen $k = 2.4$ ($\mu_\phi = 0.5\,m$). The three plots in each sub-figure correspond to \warn{$m = 0.1$}, $m = 1.0$ and \warnb{$m = 10$}, respectively, with the corresponding values \warn{$R_\hor = 0.5$}, $R_\hor = 1.1$ and \warnb{$R_\hor = 2.3$} for the horizon radius. $B(R)$ and $\phi(R)$ abruptly halts at the respective values based on the boundary conditions chosen; in particular, $\phi_{c, \nhLS} = -0.5$ for all value of the mass, as can be verified from the plot.}
	\label{figure:LBB-nh}
	\end{figure}

Since we are dealing with a special case of the non-extremal solution, connecting the near horizon form of the solution to the asymptotic form of the same is analogous to what we did in the context of the general non-extremal solution, see the discussion after equation \eq{coeffs:nh:non-ext:ord-2} in section \ref{non-extremal-soln}. In particular, by integrating the equations \eq{EEq-tt:fixed-detg:R}, \eq{EEq-xx:fixed-detg:R} and \eq{phiEOM:fixed-detg:R} from $R = R_\hor$ to $R = \infty$, we obtain \eq{Rh:non-ext:fixed-detg} and \eq{k:non-ext:fixed-detg} as the expressions for the horizon radius $R_{\hor, \nhLS}$ and $k_\nhLS$, respectively, in terms of $m$ and $\mu_\phi$. Owing to \eq{k:LBB-3} and \eq{w0-mLif:LBB-3} (by choosing the appropriate boundary conditions on the horizon) we can also express the Lifshitz mass parameter $m_\LS$ to $m$ and $\mu_\phi$
	\beq
	m_\LS = \fr{k_\nhLS}{2\rt{k_\ext}} = \fr{n + 1}{2}\lf[\fr{1 - \displaystyle{\fr{\mu_\phi}{m}}}{1 - \displaystyle{\lf(\fr{n + 1}{2n}\rf)\fr{\mu_\phi}{m}}}\rf]\fr{\hat{\ell}_\scLif}{\ell}\,.
	\eeq
As consistency demands, $m_\LS$ vanishes in the extremal limit $m \to \mu_\phi$, leaving us with the extremal solution.

Finding the solution in the bulk through numerical analysis\ft{The numerical solution in the bulk for this case has been previously obtained in \cite{lifshitz:peetetal-1} and \cite{lifshitz:peetetal-2}.} for this case is similar to the previous two cases discussed. For the present case, one important verification is whether the numerical analysis, set up for the general non-extremal solution, indeed shows the near horizon Lifshitz-Schwarzschild behaviour for the appropriate values of the parameters. We verify this in figure \ref{figure:LBB-nh}, with the following choice of boundary conditions on the horizon: $b_0 = k_\ext^{\de/2}$ and $\phi_c = \phi_0\log\rt{k_\ext}$ (see the caption of the figure for other necessary details).

\begin{figure}[h!]
	\centering
	\includegraphics[scale=0.8]{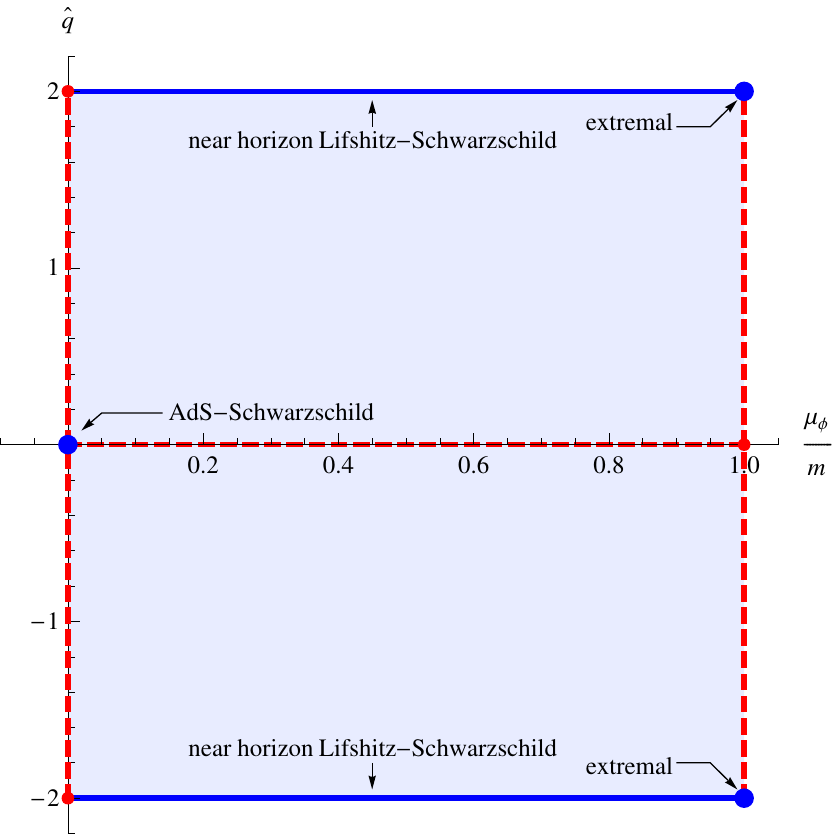}
	\caption{The parameter space of solutions. The horizontal axis corresponds to $(\mu_\phi/m)$ and ranges from $0$ to $1$, while the vertical axis corresponds to $\hatq$ and ranges between $-q_\scLif$ to $q_\scLif$. We have chosen $n = 2$ and $\al = 1$ (i.e., $q_\scLif = 2$) in order to obtain a definite number for $q_\scLif$ for the purpose of plotting, but otherwise the plot is applicable for any $n$ and $\al$. The \warn{dashed red} lines, excluding the \warnb{blue} points at $(0, 0)$ and $(1, \pm q_\scLif)$, indicate regions where there are no physical solutions. The \warnb{blue} points correspond to the special cases of the AdS-Schwarzschild solution and the extremal solutions, respectively. The \warnb{solid blue} lines at $\hatq = \pm q_\scLif$ with $(\mu_\phi/m)$ ranging between $0 < \mu_\phi/m < 1$ correspond the the near horizon Lifshitz-Schwarzschild solutions. The interiour {\color{LightBluejb}light blue} region corresponds the all the possible values of $(\mu_\phi/m, \hatq)$ for which we have general non-extremal solutions.}
	\label{figure:phase-space}
	\end{figure}

Furthermore, the numerical analysis shows\ft{To be precise, the numerical code shows  divergent behaviour of the various metric components/dilaton very near the horizon for $|\hatq| > q_\scLif$ and with different values $k$ within the bound \eq{k:non-ext:bound}.} that there are no physical solutions for the general non-extremal case, with $|\hatq| > q_\scLif$ and with different values $k$ within the bound \eq{k:non-ext:bound}, i.e., we require
	\beql{extremality-condition-2}
	|\hatq| \leqq q_\scLif\,.
	\eeq
Taken together with \eq{extremality-condition-1}, these two equations furnish two extremality conditions on the system. Thus, as expected, we have obtained non-extremal solutions when at least one of the two extremality conditions is in the form of a strict inequality and the extremal solution is only obtained when both of them are strict equalities. 
In figure \ref{figure:phase-space} we give a pictorial representation of the extremality conditions. In particular, it is clear from this plot that $\hatq$ is an independent parameter which can be varied independently of $m$ and $\mu_\phi$ (except at an isolated set of points). 

It is very interesting that  in going to non-extremality starting from the extremal solution, one always ends up on the near horizon Lifshitz-Schwarzschild solution first, after which one can decrease the value of $\hatq$ to obtain a more general non-extremal solution. This seems reasonable from the perspective of an observer near the horizon of the extremal solution. For such an observer the spacetime is Lifshitz, and the only way to overcome extremality is to throw in  matter into the brane to make the solution non-extremal, which necessarily makes the solution Lifshitz-Schwarzschild-like near the horizon.
Equivalently, there exists no smooth limit to the extremal solution starting from a general non-extremal solution i.e., for which $|\hatq| < q_\scLif$. We can take $\mu_\phi$ as close to $m$ as we want, but it is impossible to make $\mu_\phi = m$ without simultaneously making $|\hatq| = q_\scLif$. 
\section{Thermodynamics}\label{thermodynamics}
The solutions we have obtained are black objects in general relativity and hence are thermal objects governed by the laws of thermodynamics. We discuss such aspects of our solutions in this section.

Let us first consider the non-extremal solutions. Given the near horizon expressions for the metric we can compute the temperature of the black brane directly by transforming to a local Rindler coordinate and reading off the surface gravity \eq{non-ext:surface-gravity}. The temperature is then the surface gravity divided by $2\pi$\ft{The surface gravity divided by $2\pi$ is the inverse of the period of the compact Euclidean time.}, i.e.,
	\beql{T:non-ext}
	T = \fr{kR_\hor}{4\pi b_0^n\ell}\,.
	\eeq
Note that we have not set any specific boundary condition on the horizon which is why there is an explicit appearance of $b_0$ in the formula. For comparison among the various asymptocially AdS solutions, if we choose the same value for $b_0$ for all of them, then the temperature can only depend on $m$ and $\mu_\phi$ through their dependence on $k$ \eq{k:non-ext:fixed-detg} and $R_\hor$ \eq{Rh:non-ext:fixed-detg}. In particular, the temperature does not depend on $\hatq$, so that a near horizon Lifshitz-Schwarzschild can be hotter or colder than a general non-extremal solution (i.e., for which $|\hatq| < q_\scLif$). On the other hand, the extremal solution corresponds to $k = 0$ and hence its temperature is always zero. This is consistent with our expectation of an extremal solution, as well as with the fact that owing to the Lifshitz nature (scale invariance) of the extremal solution near the horizon, there cannot be a temperature (scale) associated with it. Note, that unlike the asymptotically flat dilatonic black brane solutions \cite{ghs}, the temperature \eq{T:non-ext} vanishes smoothly as extremality is approached in the limit $k\to 0$.

On the other hand, to compare the temperature of the near horizon Lifshitz-Schwarzschild solution with that of the global Lifshitz-Schwarzschild solution, we should choose the boundary condition $b_0 = k_\ext^{\de/2}$. Using \eq{T:non-ext} for non-extremal solutions along with the relations \eq{w0-mLif:LBB-3}, \eq{k:LBB-3} and \eq{hatr-hatl:LBB-3}, we obtain
	\beql{T:LBB}
	T_\nhLS = \fr{(n + z)(2m_\LS)^{\fr{z}{n + z}}R_\hor}{4\pi\ell_\scLif}\,.
	\eeq
Once we recall that the lapse functions for the near horizon Lifshitz-Schwarzschild \eq{nh-LBB} and the global Lifshitz-Schwarzschild \eq{met:LBB-3} differ by a factor of $R_\hor$ we find  perfect agreement between \eq{T:LBB} and \eq{temperature:LBB}.

Next, we obtain the  entropy density per unit $n$-volume of the general non-extremal black brane from the Bekenstein-Hawking proposal
	\beql{S:non-ext}
	S = \fr{2\pi(\text{area density per unit }n \text{-volume of the horizon})}{\kaEH^2} = \fr{2\pi b_0^nR_\hor^n}{\kaEH^2}\,.
	\eeq
Therefore, for the near horizon Lifshitz-Schwarzschild black brane, choosing the appropriate boundary condition $b_0 = k_\ext^{\de/2}$ at the horizon, and using the relations \eq{w0-mLif:LBB-3}, \eq{k:LBB-3} and \eq{hatr-hatl:LBB-3}, we obtain
	\beql{S:LBB}
	S_\nhLS = \fr{2\pi(2m_\LS)^{\fr{n}{n + z}}R_\hor^n}{\kaEH^2}\,.
	\eeq
Perfect agreement with \eq{entropy:LBB} is achieved once we recall that the transverse space \tql{area densities} in \eq{nh-LBB} and \eq{met:LBB-3} differ by a factor of $R_\hor^n$. Note that the entropy density vanishes for the extremal (near horizon Lifshitz) as well as the global Lifshitz solutions. Even though we noted earlier that in the limit $\al \to 0$, the global Lifshitz solution reduces to $\AdS_2\times\R^n$, the entropy density is a physical quantity which does not have a smooth behaviour in this limit. Stated differently, the entropy density of the near horizon Lifshitz solution is zero as long as $\al \neq 0$, but has a finite jump at $\al = 0$.  

From \eq{reln:after-integration} we can now write down Smarr's formula \cite{Smarr:1972kt} for the general non-extremal black brane
	\beql{Smarr}
	m = \fr{kR_\hor^{n + 1} + q\Phi_\hor}{2(n + 1)} = \fr{2\ell\kaEH^2TS + q\Phi_\hor}{2(n + 1)}\,.
	\eeq
Note that neither $TS$ nor the Smarr's formula itself depends on the choice of boundary condition ($b_0$) on the horizon. 
In fact, Smarr's formula can be recast in more physical terms. 
In particular, since all the solutions we have are asymptotically AdS, by AdS/CFT all such solutions correspond to thermal states in the dual CFT. The Smarr's formula \eq{Smarr} especially corresponds to the thermodynamic identity of the CFT equivalent to the well known \tql{$E = TS + pV$} for an ideal gas. By the standard rules of AdS/CFT, the temperature \eq{T:non-ext}, entropy density \eq{S:non-ext} and the mass density \eq{def:Schwarzschild-M} of the brane are identified with the temperature $T_\CFT$, entropy $S_\CFT$ and the energy $E_\CFT$, respectively, of the dual CFT. Next, the physical electric charge density \eq{def:physical-electric-charge} is identified with the number density $n_\CFT$ of the dual CFT, i.e.,
	\beq
	n_\CFT = \fr{q}{\ell\rt{2\kaEH^2}}
	\eeq    
and its conjugate chemical potential is related to the electrostatic potential with canonical dimensions \eq{redefinition}
	\beq
	\mu_\CFT = \fr{\Phi_\hor}{\rt{2\kaEH^2}}\,.
	\eeq
Putting all these together, the Smarr's formula \eq{Smarr} becomes \cite{lifshitz:peetetal-1, lifshitz:peetetal-2}
	\beql{Smarr-CFT}
	E_\CFT = \fr{n}{n + 1}\lf(T_\CFT S_\CFT + \mu_\CFT n_\CFT\rf)\,.
	\eeq
Comparison with \eq{MTS:LBB} shows 
that the global Lifshitz-Schwarzschild solution has quite different thermodynamic properties compared to the near horizon Lifshitz-Schwarzschild solution.  
\section{Summary and conclusions}\label{summary-and-conclusions}

In this paper, we have studied static, asymptotically AdS $n$-brane solutions of Einstein-Maxwell-dilaton gravity with negative cosmological constant in $(n + 2)$-dimensions where $n \geqq 2$. Depending on the values of the various parameters labeling a solution, there are five different cases to consider, as summarized in table \ref{table:list-of-solutions}. Once the boundary conditions, either at the asymptotic infinity or at the horizon, are imposed appropriately, each solution is in fact labelled by three parameters: the mass density $m$, the electric charge density $q$ and the dilatonic charge density $\mu_\phi$ carried by the brane\ft{To be precise, the physical densities with the correct dimensions are proportional to $m$, $q$ and $\mu_\phi$; see e.g., \eq{def:Schwarzschild-M} and \eq{def:physical-electric-charge}. Also, we consider densities because the actual charges, being proportional to the horizon area, are infinite due to the infinite volume of the transverse space.}. However, instead of $q$, it is actually convenient to use the parameter $\hatq$ as a label, where $\hatq$ is related to $q$ through \eq{def:k-hatq:fixed-detg}; note that the parameter $\hatq$, being a ratio of $q$ and $R_\hor^n$ \eq{Rh:non-ext:fixed-detg} is a function of the asymptotic charges only once an appropriate boundary condition is imposed.  
	\begin{table}[h!]
	\begin{tabular}{||c|c|c|c|m{8cm}||}
	\hline
	Case & $m$ & $\mu_\phi$ & $\hatq$ & Solution \\
	\hline
	\hline
	1 & $m = 0$ & $\mu_\phi = 0$ & $\hatq = 0$        & Poincar\'e patch of AdS. \\
	\hline
	2 & $m > 0$ & $\mu_\phi = 0$ & $\hatq = 0$        & AdS-Schwarzschild black brane. \\
	\hline
	3 & $m > 0$ & $\mu_\phi < m$ & $\hatq < q_\scLif$ & asymptotically AdS non-extremal dilatonic black brane. \\
	\hline
	4 & $m > 0$ & $\mu_\phi < m$ & $\hatq = q_\scLif$ & asymptotically AdS non-extremal dilatonic black brane, near horizon Lifshitz-Schwarzschild. \\
	\hline
	5 & $m > 0$ & $\mu_\phi = m$ & $\hatq = q_\scLif$ & asymptotically AdS extremal dilatonic black brane, near horizon Lifshitz. \\
	\hline
	\end{tabular}
	\caption{List of all static asymptotically AdS solutions discussed in this paper.}
	\label{table:list-of-solutions}
	\end{table}

With the exception of cases 1 and 2 in table \ref{table:list-of-solutions} (the Poincar\'e patch of AdS and the AdS-Schwarzschild black brane, respectively) all the solutions describe electrically charged black $n$-branes, and none of them admit a closed analytic form. We have therefore expressed the solutions corresponding to cases 3 to 5 as power series \eq{asymp-soln:fixed-detg:R} in the inverse power of the radial coordinate $R$ \eq{def:asymp-radial-coord-R} valid in the asymptotic ($R \gg 1$) region, as well as in a power series in the radial coordinate with respect to the horizon, $w$ \eq{def:nh-radial-coord-w}, valid near the horizon ($w \ll 1$) -- see \eq{nh-A:fixed-detg}, \eq{nh-B:fixed-detg}, \eq{nh-phi:fixed-detg} for the general forms of the near horizon solutions; we will refer to the appropriate equations for the expressions of the various exponents and other parameters as we discuss each case separately.

The central result of this paper is the general non-extremal solution (case 3, table \ref{table:list-of-solutions}) discussed in section \ref{non-extremal-soln}. This is a black brane with a regular horizon ($\met_{tt}$ vanishing linearly in $w$) and the dilaton being finite at the horizon\ft{The higher order corrections to the metric away from the horizon can be found in \eq{coeffs:nh:non-ext:ord-1} and \eq{coeffs:nh:non-ext:ord-2}.}. The asymptotic charges $m$, $q$ and $\mu_\phi$ are related to the near horizon parameters $k$ and $\hatq$ \eq{def:k-hatq:fixed-detg} through the relations \eq{k:non-ext:fixed-detg} and \eq{Rh:non-ext:fixed-detg}.

We also derived the appropriate extremality conditions satisfied by all the solutions, and they are conveniently expressed as \eq{extremality-condition-1}, \eq{extremality-condition-2}
	\beq
	m \geqq \mu_\phi; \qquad |\hatq| \leqq q_\scLif\,,
	\eeq
where $q_\scLif$ is defined in \eq{q-scale:LBB}; see figure \ref{figure:phase-space} for a pictorial representation of the extremality conditions. The first condition is in fact the actual extremality condition related to the vanishing of the surface gravity at the horizon, and this is very similar to the corresponding condition in \cite{ghs}. When the second extremality bound is saturated (i.e., $|\hatq| = q_\scLif$ but $m > \mu_\phi$), we obtain a special case of the non-extremal solution which is Lifshitz-Schwarzschild-like near the horizon (case 4, table \ref{table:list-of-solutions} -- see section \ref{nh-LBB-soln}). On the other hand, when both the bounds are saturated we obtain the extremal solution which is Lifshitz-like near the horizon (case 5, table \ref{table:list-of-solutions} -- see section \ref{extremal-soln}). These near horizon solutions are however driven away from their Lifshitz/Lifshitz-Schwarzschild-like near horizon nature towards an AdS behaviour asymptotically by the higher order corrections as given in \eq{coeffs:nh:non-ext:ord-1}, \eq{coeffs:nh:non-ext:ord-2} and \eq{coeffs:nh:ext:ord-1}, \eq{coeffs:nh:ext:ord-2}. These solutions have already been studied earlier in \cite{lifshitz:gkpt}, \cite{Chen:2010kn}, \cite{lifshitz:peetetal-1} and \cite{lifshitz:peetetal-2}.

Finally, we have also studied the thermodynamics of the solutions in section \ref{thermodynamics}. In particular, we observe that the temperature of the non-extremal solutions vanish smoothly in the extremal limit. The entropy of the extremal solution, being Lifshitz near the horizon, also vanishes. This is consistent with the third law of thermodynamics, but the behaviour is very different from the AdS-Reissner-Nordstr\"om solution. In this regard, the latter is an exact solution to this system when the dilatonic coupling is switched off, and the functional form of the near horizon Lifshitz solution reduces to $\AdS_2\times\R^n$, which is the near horizon metric of the extremal AdS-Reissner-Nordstr\"om solution.

One important issue, particularly in the light of \cite{gm-instability}, is the stability of the solutions found in this paper, against small perturbations. It would also be interesting to holographically study e.g., transport properties of the CFT duals of our solutions (especially the general non-extremal solution) with condensed matter physics applications in mind. We hope to return to these questions in the near future.

\acknowledgments 
PB would like to thank the theory group at CERN for their hospitality during various stages of this project. JB would like to thank Arnab Kundu and Robert de Mello Koch for useful discussions at the early stages of this project. DM thanks Sayan Basu for useful discussions. The work of PB and JB is supported by the NSF CAREER grant PHY-0645686 and by the University of New Hampshire through its Faculty Scholars Award Program.
\appendix
\section{Restrictions on the dilatonic potential for the existence of exact global Lifshitz solutions}\label{V-restriction}
\paragraph{Note added in version 2:} The results of this appendix has some overlap with section 1 of~\cite{Perlmutter:2010qu}. We than Eric Permutter for bringing this to our notice.

In this appendix, we prove the following no-go result: by considering the following generalization of the action \eq{our-action}
	\beql{ac:arb_V}
	\ac = \intdx{n + 2}\rt{-\det\met}\lf[\fr{\Rie}{2\kaEH^2} - V(\phi) - \fr{1}{2}(\dl\phi)^2 - \fr{e^{2\al\phi}\;\cF^2}{2}\rf]
	\eeq
we show that to obtain an exact, global Lifshitz solution of the form
	\beql{met:Lifshitz}
	A(r) = (r/\ell_\scLif)^{2z}; \qquad C(r) = (r/\ell_\scLif)^{-2}; \qquad B(r) = (r/\ell_\scLif)^2
	\eeq
$V(\phi)$ has to be constant and negative. As already noted in the introduction of this paper, the existence of global Lifshitz and Lifshitz-Schwarzschild solutions in a gravitational system makes it a candidate holographic dual of a non-relativistically scale invariant quantum field theory. Our search for possible gravitational systems involving a dilaton and a gauge field, which gives rise to an exact Lifshitz solution, is therefore conveniently narrowed down to \eq{our-action}, the one we consider in this paper\ft{Through a straight-forward generalization of the following proof, we also arrive at the same conclusion by considering an addition scalar field in \eq{ac:arb_V}, which is allowed to interact with itself as well as with the dilaton but is not coupled to the gauge field.}.

In \eq{met:Lifshitz} $z$ and $\ell_\scLif$ depend upon the various parameters of the theory provided a solution of the above form exists, as we show below. After performing the redefinition \eq{redefinition} to get rid of factors of $\kaEH^2$, the equations of motion from \eq{ac:arb_V} take the following form under the same assumptions as those in section \ref{stat-Rniso-met-EOM} ($\ell$ is an arbitrary length scale at this point)
	\bseql{EEq:arb_V}
	\beql{EEq-tt:arb_V}
	\Dl^2\log A(r) = \fr{n-1}{n}\fr{q^2}{\ell^2}\fr{e^{-2\al\phi(r)}}{B(r)^n} - \fr{2}{n}V(\phi)
	\eeq

	\beql{EEq-tr:arb_V}
	\fr{n}{4}\fr{B'(r)}{B(r)}\lf[\fr{A'(r)}{A(r)} + \fr{C'(r)}{C(r)} - \fr{B'(r)}{B(r)}\rf] - \fr{n}{2}\lf[\fr{\ed}{\dr}\fr{B'(r)}{B(r)}\rf] = \fr{\phi'(r)^2}{2}
	\eeq

	\beql{EEq-xx:arb_V}
	\Dl^2\log B(r) = -\fr{1}{n}\fr{q^2}{\ell^2}\fr{e^{-2\al\phi(r)}}{B(r)^n} - \fr{2}{n}V(\phi)
	\eeq
and the dilaton equations of motion is
	\beql{phiEOM:arb_V}
	\Dl^2\phi(r)  = -\fr{\al q^2}{\ell^2}\fr{e^{-2\al\phi(r)}}{B(r)^n} + \fr{\ed V}{\dphi}
	\eeq
	\eseq
To begin with, note that the Laplacian of any function $\psi(r)$, for the metric \eq{met:Lifshitz} is
	\beq
	\Dl^2\psi(r) = \fr{r^2}{\ell_\scLif^2}\lf[\psi''(r) + \fr{\psi'(r)}{r}(z + n + 1)\rf]
	\eeq
Furthermore, if $\psi(r)$ has a logarithmic behaviour, i.e., $\psi(r) = \psi_c + \psi_0\log (r/\ell_\scLif)$ where $\psi_c$ and $\psi_0$ are constants, then
	\beql{Laplacian:Lifshitz}
	\Dl^2\psi(r) = \fr{\psi_0}{\ell_\scLif^2}(z + n)
	\eeq
Subtracting \eq{EEq-xx:arb_V} from \eq{EEq-tt:arb_V} and using the above expression for the Laplacian on the left hand side
	\beql{EEq-tx:Lifshitz}
	\Dl^2\log\lf[\fr{A(r)}{B(r)}\rf] = \fr{2(z - 1)(z + n)}{\ell_\scLif^2} = \fr{q^2}{\ell^2}\;e^{-2\al\phi(r)}\;(r/\ell_\scLif)^{-2n}
	\eeq
Since the left hand side (term in the middle) is constant, consistency demands $e^{-2\al\phi(r)}\;(r/\ell_\scLif)^{-2n}$ is a constant as well. Writing this constant as $e^{-2\al\phi_c}$, where $\phi_c$ is another real constant, we have
	\beql{phi:Lifshitz}
	\phi(r) = \phi_c - \fr{n}{\al}\log (r/\ell_\scLif)
	\eeq
This also means that the dilatonic coupling terms in \eq{EEq:arb_V} are constant
	\beql{dilatonic-coupling:Lifshitz}
	\fr{q^2}{\ell^2}\fr{e^{-2\al\phi(r)}}{B(r)^n} = \fr{\hatq^2}{\ell_\scLif^2}; \qquad \hatq = \fr{q\ell_\scLif e^{-\al\phi_c}}{\ell}
	\eeq
Getting back to \eq {EEq-tx:Lifshitz}, we find $\hatq$ to be restricted as follows
	\beql{fixed-hatq:Lifshitz}
	\hatq^2 = 2(z - 1)(n + z)
	\eeq
Next, if we use the ansatz for the metric components \eq{met:Lifshitz} and the logarithmic behaviour of the dilaton just derived in \eq{phi:Lifshitz} into \eq{EEq-tr:arb_V}, we find that the scaling exponent is given by
	\beql{def:z:Lifshitz}
	z = 1 + \fr{n}{2\al^2}
	\eeq
Now, using \eq{Laplacian:Lifshitz}, \eq{fixed-hatq:Lifshitz} and \eq{def:z:Lifshitz} in \eq{phiEOM:arb_V}, we can show that the radial variation of the potential vanishes as well
	\beq
	\fr{\ed V}{\dphi} = \Dl^2\phi(r) + \fr{\al\hatq^2}{\ell_\scLif^2} = 0 \quad\im\quad \fr{\ed V}{\dr} = \fr{\ed V}{\dphi}\phi'(r) = 0
	\eeq
Thus the only potential consistent with a scaling solution of the form \eq{met:Lifshitz} is a constant. We are now going to fix the arbitrary length scale $\ell$ by setting it equal to the scale of the constant potential/vacuum energy as follows
	\beql{Vphi-restricted:Lifshitz}
	V(\phi) = \fr{kn(n + 1)}{\ell^2}
	\eeq
where $k$ is either $-1$ or $0$ or $1$. Using \eq{EEq-xx:arb_V} now, we find
	\beq
	\hatq^2 = -\fr{2kn(n + 1)}{2\al^2 + 1}\fr{\ell_\scLif^2}{\ell^2}
	\eeq 
Since $\hatq$ is real, we are forced to choose $k = -1$. From \eq{dilatonic-coupling:Lifshitz} The charge $q$ is then given by
	\beql{q:Lifshitz}
	q^2 = \fr{2n(n+1)}{2\al^2 + 1}
	\eeq
where, the constant $\phi_c$, which remains unconstrained up to this point, has been set to zero without any loss in generality. Note that $q$ is independent of both $\ell_\scLif$ and $\ell$. Finally, to satisfy \eq{EEq-xx:arb_V} (and \eq{EEq-tt:arb_V}) the constant $\ell_\scLif$ needs to be fixed at
	\beql{scale:Lifshitz}
	\ell_\scLif = \ell\rt{\fr{(n + z - 1)(n + z)}{n(n + 1)}} = \ell\rt{\fr{(1 + 2\al^2)(n + 2(n + 1)\al^2)}{2\al^4(n + 1)}}
	\eeq
We can regard $\ell_\scLif$ as the length scale associated with the Lifshitz geometry (like $\ell$ in $\AdS_{n + 2}$) and call it the Lifshitz scale. Note that $\ell_\scLif \geqq \ell$ in this particular solution. Also, $\ell_\scLif \to \ell$ as $\al \to \infty$ i.e., as  $z \to 1$, and the scaling solution becomes identical to the Poincar\'e patch of $\AdS_{n + 2}$ with $\ell$ as its radius.
\section{The Hawking-Horowitz Mass}\label{appendix:mass}
In this appendix, we calculate the mass of $n$-brane solutions in asymptotically AdS and Lifshitz spacetimes via the prescription of \cite{haw-hor}. As with any Hamiltonian formulation, we first need to consider a foliation of the spacetime by constant time slices $\Sg_t$. In our case, we can choose the coordinate time $t$ as the global time function, and parametrize the time slices (which are also the Cauchy surfaces) $\Sg_t$ with $t$. The unit timelike normal on $\Sg_t$ is 
	\beq
	(\hatn_t)_\mu = -\rt{A(r)}\dl_\mu t = -\rt{A(r)}\{1, 0, ..., 0\} \qquad\iim\qquad (\hatn_t)^\mu = \fr{1}{\rt{A(r)}}\{1, 0, ..., 0\}
	\eeq
Note that the timelike Killing vector $(\chi_t)^\mu = \{1, 0, ..., 0\}$ is related to the normal vector through
	\beql{appendix:reln:chi_t-n}
	\chi_t = \rt{A(r)}\,\hatn_t
	\eeq
Let us denote by $\hmet(\Sg_t)_{\mu\nu}$ the (spatial) metric induced by $\met_{\mu\nu}$ on $\Sg_t$. The induced metric is related to the metric through
	\beq
	\hmet(\Sg_t)_{\mu\nu} = \met_{\mu\nu} + (\hatn_t)_\mu(\hatn_t)_\nu = \diag\lf\{0, C(r), B(r), ..., B(r)\rf\}
	\eeq
The above construction allows us to obtain the lapse function and the shift vector. Let $t^\mu$ (not to be confused with the coordinate time $t$) be the vector field generating time translations and satisfying $t^\mu\Dl_\mu t = 1$. The natural choice for a $t^\mu$ is the timelike Killing vector itself
	\beq
	t^\mu = (\chi_t)^\mu = \{1, 0, ..., 0\}
	\eeq
The lapse function $N$ and the shift vector $N^\mu$ are defined in the usual way \cite{gr:books} as the decomposition of $t^\mu$
	\beq
	t^\mu = N(\hatn_t)^\mu + N^\mu
	\eeq
We can read off the lapse and the shift from \eq{appendix:reln:chi_t-n} 
	\beql{appendix:lapse-shift}
	N = \rt{A(r)}; \qquad N^\mu = 0
	\eeq
We also have a \tql{boundary near infinity}, $\Sg_\infty$, on which we specify the asymptotic behaviour of the various fields. To obtain $\Sg_\infty$, consider a foliation of the spacetime through the hypersurfaces $\Sg_r$, which are slices of the spacetime at fixed values of the radial coordinate $r$\ft{We have used $r$ itself to parametrize the hypersurfaces $\Sg_r$}. We can then identify $\Sg_\infty$ with the asymptotic boundary of the spacetime located at $r = \infty$, i.e., $\Sg_\infty \equiv \Sg_{r = \infty}$. The unit spacelike normal on $\Sg_r$, denoted by $\hatn_r$, is given by
	\beq
	(\hatn_r)_\mu = \rt{C(r)}\dl_\mu r = \rt{C(r)}\{0, 1, 0, ..., 0\} \qquad\iim\qquad (\hatn_r)^\mu = \fr{1}{\rt{C(r)}}\{0, 1, 0, ..., 0\}
	\eeq  
The fact that $\hatn_t\cdot\hatn_r = 0$ clearly shows that $\hatn_r$ is tangential to $\Sg_t$ while $\hatn_t$ is tangential to $\Sg_r$. The induced metric on $\Sg_r$, to be denote by $\hmet(\Sg_r)_{\mu\nu}$, is
	\beq
	\hmet(\Sg_r)_{\mu\nu} = \met_{\mu\nu} - (\hatn_r)_\mu(\hatn_r)_\nu = \diag\lf\{-A(r), 0, B(r), ..., B(r)\rf\}
	\eeq
The hypersurfaces $\Sg_t$ and $\Sg_r$ are themselves foliated by the $n$ dimensional transverse spaces $\Sg_{t, r}$ with induced metric
	\beq
	\begin{split}
	\hmet(\Sg_{t, r})_{\mu\nu} & = \hmet(\Sg_t)_{\mu\nu} - (\hatn_r)_\mu(\hatn_r)_\nu = \hmet(\Sg_r)_{\mu\nu} + (\hatn_t)_\mu(\hatn_t)_\nu \\
	& = \met_{\mu\nu} + (\hatn_t)_\mu(\hatn_t)_\nu - (\hatn_r)_\mu(\hatn_r)_\nu = \diag\lf\{0, 0, B(r), ..., B(r)\rf\}
	\end{split}
	\eeq
The final expression for the energy involves the trace of the extrinsic curvature of $\Sg_{t, r}$ due to its embedding in $\Sg_t$ evaluated at $r = \infty$. For a finite value of $r$, the extrinsic curvature is given by\ft{The extrinsic curvature of $\Sg_{t, r}$ due to its embedding in $\Sg_r$ vanishes.}
	\beq
	\ExtK(\Sg_{t, r})_{\mu\nu} = \fr{1}{2}\LieD_{\hatn_r}\hmet(\Sg_{t, r})_{\mu\nu} = \fr{1}{2}\LieD_{\hatn_r}\met_{\mu\nu} + \fr{1}{2}\LieD_{\hatn_r}(\hatn_t)_\mu(\hatn_t)_\nu - \fr{1}{2}\LieD_{\hatn_r}(\hatn_r)_\mu(\hatn_r)_\nu
	\eeq
The trace of the this extrinsic curvature is then
	\beql{appendix:trK}
	\ExtK(\Sg_{t, r}) = \met^{\mu\nu}\ExtK(\Sg_{t, r})_{\mu\nu} = \fr{nB'(r)}{2B(r)\rt{C(r)}}
	\eeq
The Hamiltonian for our case is now given by
	\beql{appendix:Hamiltonian}
	\Ham = \fr{1}{2\kaEH^2}\int\limits_{\Sg_t}NH_\text{con} - \fr{1}{\kaEH^2}\int\limits_{\Sg_{t, r = \infty}} N\ExtK(\Sg_{t, r})
	\eeq
where $N$ is the lapse \eq{appendix:lapse-shift}, $H_\text{con}$ is the Hamiltonian constraint and we have dropped terms involving the shift vector since it vanishes for our case \eq{appendix:lapse-shift} (see \cite{haw-hor} for the complete expression). Also, as emphasized in \cite{haw-hor}, in case of spatially non-compact geometries (as with flat $n$-branes) the Hamiltonian might be divergent when evaluated on a solution. In such cases, one needs to consider a reference background which is a static solution of the field equations, such that, the reference background is asymptotically approached by any physical solution whose energy we want to compute. The physical Hamiltonian is then obtained by subtracting the Hamiltonian evaluated on the reference background from the usual Hamiltonian \eq{appendix:Hamiltonian}, and the energy of a given solution is the value of the physical Hamiltonian evaluated on the solution 
	\beql{appendix:energy-gen}
	E = -\fr{1}{\kaEH^2}\int\limits_{\Sg_\infty} N\lf[\ExtK(\Sg_{t, r}) - \ExtK_0(\Sg_{t, r})\rf]_{r=\infty}
	\eeq
where we have used the fact that that $H_\text{con}$ vanishes on a physical solution and have denoted the parts evaluated on the reference background by a subscript $0$. It should be emphasized that for the above prescription to work, it is crucial for the lapse of the solution of interest and the background solution to agree asymptotically (which allowed us to factor the same out from the expression inside the integral). The square bracketed quantity inside the integral above secretly contains the volume factor of the transverse space.

For the special class of asymptotically AdS solutions that we studied in this paper, we must choose the reference background to be the Poincar\'e patch of the pure $\AdS_{n+2}$ solution with a constant dilaton. However, we need to remember that the integral over $\Sg_\infty$ gives an infinite contribution owing to the infinite $n$-volume of the flat transverse space. Therefore, the physical quantity to work with is the energy density per unit $n$-volume, obtained through dividing the energy computed from \eq{appendix:energy-gen} by the volume of the transverse space. In what follows,  we denote by $E$ this energy density (as opposed to the infinite total energy) with the hope that there will be no confusion after our clarifying remarks above. Now, all the solutions discussed in this paper, including the special cases of the AdS-Schwarzschild and the AdS-Reissner-Nordstr\"om solutions as well as the various cases of the dilatonic black brane solutions, satisfy the metric fixing condition \eq{def:fixed-detg:R}. We can then employ this condition to eliminate the function $C(R)$. Furthermore, all such solutions have a series expansion in $1/R$ analogous to \eq{asymp-soln:fixed-detg:R}, where for the AdS-Schwarzschild and the AdS-Reissner-Nordstr\"om solutions the series truncate after a finite number of terms. Using the explicit expression \eq{appendix:trK} for the trace of the extrinsic curvature in \eq{appendix:energy-gen} the energy is
	\beql{appendix:energy-asymp-AdS}
	E = \fr{nm}{\kaEH^2\ell}
	\eeq
Similarly, to find the energy of the global Lifshitz-Schwarzschild solution discussed in section \ref{L-LBB}, we use the global Lifshitz solution as the reference background. Following the same procedure as discussed above in the context of asymptotically AdS solutions, we employ \eq{appendix:energy-gen} to obtain \eq{M:LBB}. 

\end{document}